\documentclass[%
 aip,
 amsmath,amssymb,
 reprint,%
]{revtex4-1}

\usepackage[utf8]{inputenc}

\usepackage{braket}
\usepackage{xcolor}
\usepackage{amsmath}
\usepackage{graphicx}
\usepackage{subfig}
\usepackage{caption}
\usepackage{placeins}

\begin{document}

\preprint{AIP/123-QED}

\title{An excitation matched local correlation
approach to\\excited state specific perturbation theory}

\author{Rachel Clune}
\affiliation{
Department of Chemistry, University of California, Berkeley, CA 94720
}

\author{Eric Neuscamman}
\email{eneuscamman@berkeley.edu}
\affiliation{
Department of Chemistry, University of California, Berkeley, CA 94720 
}
\affiliation{Chemical Sciences Division, Lawrence Berkeley Nat.\ Lab, Berkeley, CA, 94720}

\date{\today}

\begin{abstract}
We develop a cubic scaling approach to excited-state-specific second
order perturbation theory in which the completeness of a
local correlation treatment is carefully matched between the
ground and excited state.
With this matching, the accuracy of the parent method is
maintained even as substantial portions of the correlation energy
are neglected.
Even when treating a long-range charge transfer excitation,
cubic scaling is achieved in systems with as few as ten non-hydrogen atoms.
In a test on the influence of an explicit solvent molecule on a
long range charge transfer, the approach is qualitatively more accurate
than EOM-CCSD and reproduces CC3's excitation
energies and excited state potential energy surface to within about
0.1 eV and 0.5 kcal/mol, respectively.
\end{abstract}

\maketitle

\section{Introduction}
\label{sec:intro}

Recent advances in excited-state-specific wave function
correlation methods
\cite{zheng2019performance,lee2019excited,N5-ESMP2,tuckman2024aufbau,tuckman2025improving}
have shown significant promise for improving the accuracy
of excitation energy predictions, especially in challenging
cases such as charge transfer.
However, like most wave function approaches to electron
correlation, the usefulness of these methods is constrained
by the rapid growth of their computational costs with
system size.
One general approach to combating such costs
is to employ local correlation treatments,
\cite{
pulay1983localizability,
pulay1986force,
pulay1986orbital,
saebo1987fourth,
saebo1993efficient,
hampel1996local,
maslen1998noniterative,
schutz1999low-I,
schutz2000low-III,
schutz2001low-IV,
schutz2002low-V,
werner2003fast,
df_lcc2,
lt_df_lcc2,
LPNO-CCSD,
DLPNO-CCSD,
neese2009efficient,
auer2006dynamically,
russ2008local,
subotnik2005local,
subotnik2008exploring,
yang2012orbital,
sparta2014chemical,
kats2014speeding,
pno_ms_caspt2,
liakos2015exploring,
riplinger2016sparse,
tew2019principal,
guo2020linear,
wang2023local,
shi2024local,
nagy2024state,
yang2024making,
russ2004local,
liang2025efficient}
and, thanks to the strong parallels between
excited-state-specific wave function methods and their
ground state counterparts, many ground state local correlation
treatments should readily generalize to work with
excited-state-specific methods.
In considering such an approach in the context of predicting
excitation energies, however, one quickly realizes that the
usual local correlation goal of recovering a high fraction
of the correlation energy
(e.g.\ the 99.999\% sought in one recent
\cite{wang2023local} approach)
will involve much unnecessary effort.
In particular, when taking the energy difference that produces
an excitation energy, many of the contributions to the ground
and excited state correlation energies, such as those from
core electrons, are basically going to cancel out.
This observation leads to some interesting questions.
Can we exploit this effect to predict accurate excitation energies while
intentionally neglecting substantial chunks of the correlation energy?
If so, could this approach facilitate an early crossover to the
low scaling regime promised by local correlation methods?
In exploring these questions in the context of an
excited-state-specific perturbation theory, we find that
the answers are yes and yes.

Intentionally forgoing specific parts of the
correlation treatment that are expected to cancel out in
energy differences is a widely practiced approach in
electronic structure theory.
Perhaps the most ubiquitous examples are the frozen core
\cite{baerends1973self}
and pseudopotential \cite{STU, BFD, cECP, ccECP}
approximations, which exploit the fact that correlation effects for
core electrons are mostly insensitive to chemical changes.
Another example is
difference dedicated configuration interaction,
\cite{miralles1993specific,garcia1995iterative,chien2017recovering}
in which the determinant expansion is limited to excitations
that have at least one active orbital.
This strategy explicitly excludes correlation between most pairs
of electrons from the configuration interaction treatment
on the theory that those correlation contributions are insensitive
to the chemical change.
In some applications, \cite{garcia1995iterative}
the effects of this insensitive part of the correlation
is estimated using M{\o}ller–Plesset perturbation theory (MP2).
In the present study, we will pursue an approach that is similar in
spirit, but we will use an excited state perturbation theory as
the ``high level'' method and a local correlation
approximation matched to the specific excitation in question
to treat the less sensitive correlation contributions.

Although in principle a number of different local correlation
approaches could be employed, we use
pair natural orbitals (PNOs) in this study due to their
well established track record and efficient realization
through domain-based local PNO (DLPNO) methods.
\cite{sparta2014chemical}
In the DLPNO approach, a low cost perturbation
theory is carried out for each pair of localized 
occupied orbitals in a limited virtual space consisting
of a domain of projected atomic orbitals in the vicinity
of that occupied pair.
A small set of the most significant natural orbitals from
this perturbation theory are retained as the PNO
virtual space for that pair, dramatically reducing the
number of virtual orbitals that enter into subsequent
full-system correlation treatments. \cite{DLPNO-CCSD}
In other words, the key PNO preparation step in the DLPNO
approach boils down to a careful analysis of the localized
molecular orbitals from the mean-field reference state.
Although this analysis is usually carried out for
ground state orbitals, the rise of excited-state-specific
mean field references
\cite{gilbert2008self,besley2009self,ESMF,GVP,hardikar2020self,hait2020excited,carter2020state}
provides an opportunity to do so for excited states as well.
In this study, we aim to go one step further by
constructing an excitation matched local
correlation (EMLC) treatment in which
correlation contributions are
separated into those that involve orbitals
strongly affected by the excitation and those
that do not.
The former will be treated as they were in the original theory,  while the
latter, as they are expected to be insensitive to the
excitation and to thus essentially cancel
out in the excitation energy difference, will be
treated with an unusually aggressive 
DLPNO approach (i.e.\ with a very loose threshold).
So long as these latter parts are indeed well matched
between the states so that the error cancels,
the resulting cost reduction should come without
degrading excitation energy accuracy.

Although the EMLC strategy is not inherently specific to any
one correlation theory, this study explores its application
to our recently developed excited-state-specific
second order perturbation theory (ESMP2).
\cite{N5-ESMP2}
Much as ground state second order
M{\o}ller–Plesset perturbation (MP2) theory
results from applying Rayleigh–Schr{\"o}dinger
theory atop a restricted Hartree Fock (RHF)
reference, \cite{helgaker2013molecular}
ESMP2 is built atop an excited state mean field
(ESMF) reference. \cite{GVP}
This starting point makes the EMLC strategy
relatively straightforward:
the degree to which each molecular orbital is
involved in the excitation and the degree to which
its shape has changed relative to the ground state
can be readily analyzed via comparisons between
the ESMF and RHF wave functions.
As we will see in the results, exploiting this knowledge
allows the low scaling regime to be reached in
remarkably small system sizes without
compromising accuracy.

\section{Theory\label{sec:theory}}

\subsection{Overview}

Like many incarnations of Rayleigh–Schr\"{o}dinger perturbation theory,
MP2 and ESMP2 can be organized around a linear equation
and a second order energy expression
\begin{align}
    \label{eqn:lin-eq}
    \left( \mathbf{H}^{(0)} - E^{(0)} \right) \vec{t} &= -\hspace{0.5mm}\vec{r} \\
    \label{eqn:2nd-energy}
    E^{(2)} &= \vec{t} \cdot \vec{r}
\end{align}
in which $E^{(0)}$ is the zeroth order energy and $\mathbf{H}^{(0)}$ is the matrix
representation
\begin{align}
H^{(0)}_{\nu\mu} = \langle \nu|\hat{H}^{(0)}|\mu\rangle
\end{align}
of the zeroth order Hamiltonian in the basis of many-body states $|\mu\rangle$
whose span forms the first order interacting space $\Omega$ in which the theory plays out.
The elements of the right-hand side vector $\vec{r}$ are the first order
Hamiltonian's matrix elements between the reference state $|0\rangle$ and each
of the basis vectors.
\begin{align}
\label{eqn:rhs-elems}
r_\mu = \langle\mu|\hat{H}^{(1)}|0\rangle
\end{align}
In MP2, the reference is RHF, while in ESMP2 it is ESMF.
Finally, the vector $\vec{t}$ contains the expansion coefficients, typically referred
to as amplitudes, for the first order correction to the wave function.
\begin{align}
|\Psi^{(1)}\rangle=\sum_\mu t_\mu |\mu\rangle
\end{align}
In MP2 theory, $\Omega$ is the span of all doubly excited determinants, while
in ESMP2 it is the span of these plus the small slice of triply excited determinants
that are two excitations away from the primary determinants
of the ESMF reference. \cite{N5-ESMP2}

In the present study, portions of $\Omega$ will be replaced by small
sets of PNO-based doubly and triply excited determinants so as to prevent the cost of
working with Eqs.\ (\ref{eqn:lin-eq}) and (\ref{eqn:2nd-energy}) from growing
faster than cubically with system size $N$.
Note that, although this study achieves this goal with PNOs,
we expect that the basic idea should be compatible with
other approaches to local correlation as well.
To get to cubic scaling, we must address both the cost of
evaluating the right-hand side vector $\vec{r}$ and the cost of
solving the linear equation, after which the cost of evaluating
the 2$^{\mathrm{nd}}$ order energy will follow automatically.
In canonical MP2, thanks to the fact that its
$\mathbf{H}^{(0)}$ is the diagonal Fock matrix, we are starting
from an $O(N^4)$ cost for the linear equation but an $O(N^5)$ cost for the
right-hand side due to its reliance on key
two-electron integral transformations. \cite{helgaker2013molecular}
In ESMP2, similar integral transformations also give the right hand side
an $O(N^5)$ cost, but the overall situation is a bit more challenging
because the the linear equation is not diagonal.
In practice, it is solved via the conjugate gradient method, whose bottleneck is the
action of $\mathbf{H}^{(0)}$ on an amplitude vector,
which also has an $O(N^5)$ cost. \cite{N5-ESMP2}
Thus, to achieve this study's goal of $O(N^3)$ for EMLC-ESMP2, both evaluating
the right-hand side and performing the linear transformation must be made less expensive.
Central to this goal is a careful definition of how we use PNOs
to pare down the first order interacting space $\Omega$.

\subsection{First Order Interacting Space}
\label{sec:first-order-space}

In EMLC-ESMP2, we reduce cost by identifying and removing those portions of MP2's and ESMP2's
first order interacting spaces whose effects on correlation energies are expected to
cancel out in the subtraction that yields the excitation energy.
In the present study, the strategy will be to carefully demarcate and leave unchanged
any correlation contributions associated with orbitals involved in or strongly affected by the excitation
and then to apply PNO-based approximations to everything else.
If we take the $n\rightarrow\pi^*$ excitation in propanamide as an example,
we might reasonably expect that correlation associated with the amide group's
electrons should be treated carefully, but that correlation in the methyl cap's
CH bonds is likely not affected much by the excitation and so could be safely approximated.
Towards this end, we will separate the occupied and virtual orbitals into ``active''
and ``inactive'' categories based on the level of their involvement in the excitation.
Roughly speaking, active orbitals are those that significantly participate in the excitation,
as measured by the singular values of the ESMF
configuration interaction singles (CIS) coefficient matrix.
For now, acknowledging this active-inactive distinction is enough for us to be
able to demarcate which parts of the interacting space $\Omega$ will receive
local correlation approximations and which will not.
In Section \ref{sec:comp-detail} below, we will flesh out the
details of how orbitals are divided into active and inactive sets.

Once we have chosen active and inactive orbitals,
we divide the amplitudes in $\vec{t}$
and their corresponding basis vectors into different blocks,
some of which will be left as is and some of which will have their virtual space
approximated by PNOs.
Let's start with MP2 where, for each of the four indices on an individual amplitude
$t_{ij}^{ab}$, that index could be either active or inactive.
Taking all $2^4=16$ possible cases, we will have one block in which all four indices
are inactive and 15 blocks in which one or more is active.
Assuming that active orbitals' correlation could be strongly affected by the excitation,
we retain all amplitudes within the blocks with one or more active indices,
only applying the PNO approximation to the one quadruply-inactive block.
That block gets split into pieces, with each pair
of (localized) inactive occupied orbitals
being given its own tailored set of virtual orbitals.
As we will discuss in the next section, the choice of orbitals for this block is
made so as to make essentially the same PNO approximation in both the ground and
excited states.
As we expect each inactive occupied pair to have $O(1)$ PNOs,
we end up with $O(N^2)$ MP2 amplitudes in the quadruply inactive block.
All told, the other 15 blocks contain $O(N^3)$ amplitudes,
as they each have at least one active index,
and we expect there to be only $O(1)$ active orbitals,
as a single electron excitation typically only strongly affects the
orbitals near its particle and hole sites. Note that this is even true
for a long range CT excitation in which the particle and hole are
far apart from each other.

For ESMP2, the strategy for the doubles is the same, but we must also
address the triples.
To begin, we limit our focus to those triples that contain at least one
transition orbital pair (TOP) whose indices are active.
The TOPs are the occupied-virtual orbital pairs that result from a
singular value decomposition of the ESMF configuration interaction matrix.
\cite{N5-ESMP2}
As there are only $O(1)$ active orbitals,
we thus begin with only $O(N^4)$ triples.
We then proceed in the same way as we did for the doubles:
we separate these triples out into blocks based on
whether each of their four remaining indices is active or inactive.
As in the doubles, we retain all amplitudes in blocks where any of
these remaining indices is active,
while we apply a PNO treatment to the block where all four of these
remaining indices are inactive.
Specifically, for each pair of (localized) inactive occupied orbitals
in this block,
the two inactive virtual indices are restricted to the PNOs
that were used for that inactive pair in the corresponding doubles.
As with MP2, this leaves us with $O(N^3)$ amplitudes in the blocks in which all
amplitudes were retained
and $O(N^2)$ amplitudes in the PNO-approximated quadruply-inactive blocks.
Thus, all told, we have reduced the number of amplitudes in the theory
from $O(N^4)$ to $O(N^3)$, which is a clear prerequisite for cubic scaling.
Most blocks of the MP2 and ESMP2 amplitudes have been left alone, but those
with four inactive indices have been downsized by replacing the full inactive
virtual space with PNO-based virtual spaces tailored to the needs of each
inactive occupied pair.

To minimize the cost of working with this downsized set of triples,
we only employ the full off-diagonal $\mathbf{H}^{(0)}$ amongst triples
in which the active TOP corresponds to one of the large ``primary''
configuration state functions (CSFs) in the ESMF state
(see Section \ref{sec:comp-detail} for how primary is defined).
For triples whose active TOP is non-primary, we discard
the off-diagonal elements of $\mathbf{H}^{(0)}$, thus decoupling them
from the other amplitudes \cite{N5-ESMP2}
and simplifying the job of
solving Eq.\ (\ref{eqn:lin-eq}).
Note that doubles amplitudes employ the full off-diagonal $\mathbf{H}^{(0)}$
amongst themselves and between themselves and the primary-TOP triples.
Finally, working in the TOP basis,
we truncate the initial ESMF wave function by dropping the
CSFs with the smallest singular values
(see Section \ref{sec:comp-detail}).
To mitigate the effect of this truncation, we include a crude
estimate for the corresponding second order energy contribution.
Specifically, we scale up the correlation
energy contribution from the triples treated
with the diagonal $\mathbf{H}^{(0)}$ approximation
by the ratio of the square weight sums of the
non-primary ESMF CSFs before and after we apply the truncation.
This approach is essentially assuming that the truncated
CSFs would have contributed the same correlation energy,
weight for weight, as the active but non-primary CSFs that
were retained.

\subsection{Excitation Matched Orbitals}

The choice of the molecular orbital basis is central to the
setup of EMLC-ESMP2.
The idea is to only localize and subject to local correlation approximations
orbitals that are not strongly involved in or influenced by the excitation,
and to ensure that the local correlation approximations we do make
are well balanced between the ground and excited state.
By only localizing the inactive occupied orbitals that are little affected by
the excitation, it should be possible to ensure that each ground state
(RHF) localized orbital has a closely matching partner among the
excited state (ESMF) localized orbitals.
This way, when the virtual space for a given pair of localized occupied
orbitals is truncated into a small set of PNOs, the approximation that
this induces in the ground state will be matched by a nearly identical
approximation in the excited state, allowing for effective error cancellation
in the energy difference that forms the excitation energy.
The idea is that, thanks to this careful cancellation and the fact that we
only apply the local correlation approximation to orbitals
not strongly involved in the excitation,
it should be possible to be more aggressive than usual in the virtual space
truncation without meaningfully affecting excitation energies.

To achieve this orbital setup, we begin by transforming the ESMF state
into the transition orbital pair (TOP) basis, \cite{N5-ESMP2}
in which each occupied orbital excites into its own unique virtual
orbital and the excitation weights can be sorted from largest to smallest.
To determine which occupied orbitals should be localized and which should
be left alone, we first project each ESMF occupied TOP orbital into the
RHF occupied orbital space and measure what fraction ($\mu_i$ for the $i$th
occupied TOP orbital)
is deleted by the projection.
This serves as a measure for how much the orbital's shape changed
during ESMF's post-excitation orbital relaxation.
We can also measure the orbital's involvement in the excitation by
inspecting the corresponding TOP singular value $\sigma_i$, which is the
CIS coefficient in the TOP basis for this occupied orbital.
If both $\mu_i$ and $\sigma_i$ are small enough, then we say that the
orbital is not strongly involved in or affected by the excitation, and
we flag $i$ for possible localization.
We then repeat the same test on the virtual TOP orbitals and take our
final list $\Xi$ of indices of to-be-localized occupied orbitals
as all those that were flagged during the occupied orbital analysis
or had their TOP partners flagged during the virtual orbital analysis.
At this point, we are ready to perform occupied-occupied rotations among
the orbitals in $\Xi$ to convert them to excitation matched localized orbitals.

This localization begins by evaluating the orbital overlaps $S_{ij}$ between the
RHF and the ESMF orbitals to be localized,
\begin{align}
    \label{eqn:rhf_esmf_tbl_overlap}
    S_{ij} = \langle \phi_i | \theta_j \rangle
    \qquad\qquad i\in\Xi,j\in\Xi,
\end{align}
where $\phi$ and $\theta$ represent RHF and ESMF orbitals, respectively.
To get ground state inactive occupieds that closely match their excited
state counterparts, we use these overlaps to approximately reconstruct
the to-be-localized ESMF orbitals as linear combinations of
the to-be-localized RHF orbitals,
\begin{align}
    \label{eqn:esmf_orbs_expressed_via_rhf_orbs}
    | \Tilde{\theta}^0_j \rangle = \sum_i S_{ij} |\phi_i\rangle
    \qquad\qquad i\in\Xi,j\in\Xi,
\end{align}
after which we orthonormalize these orbitals $\Tilde{\theta}^0_j$ amongst
each other via symmetric L\"{o}wdin orthonormalization
\cite{lowdin1950non}
to produce the orbitals $\Tilde{\theta}_j$ that span the space
of to-be-localized RHF orbitals and closely approximate the to-be-localized
ESMF orbitals.
A modified Pipek-Mezey procedure (see Section \ref{sec:comp-detail}) is
then performed to find the unitary mixing of the orbitals $\Tilde{\theta}_j$
that best localizes them.
We apply this unitary transformation to the orbitals $\Tilde{\theta}_j$
to produce a set of localized RHF orbitals,
\textit{and we apply exactly the same unitary transformation}
to the orbitals $\theta_j$ to produce a set of localized ESMF orbitals that
are, thanks to our careful obital matching, nearly identical to the
localized RHF orbitals.
At this point, we are ready to apply a DLPNO treatment in which the
truncations of occupied pairs' virtual spaces have closely matched
meanings in the ground and excited states.

\subsection{Cubic scaling}
\label{sec:cubic}

With the number of amplitudes growing cubically with system size,
it remains to show that the cost of solving for the amplitudes
grows cubically as well.
Two tasks are required: evaluating the vector $\vec{r}$ on the
right hand side of Eq.\ (\ref{eqn:lin-eq}) and inverting that
equation to solve for the amplitudes $\vec{t}$.
Let us first consider the inversion of the linear equation, which
in practice is done iteratively using the conjugate gradient algorithm.
\cite{templates-linear-systems}
The key step is therefore the application of the matrix $\mathbf{H}^{(0)}$
to vectors like $\vec{r}$.
For the portions of the space where we are approximating
$\mathbf{H}^{(0)}$ as diagonal, this step is already $O(N^3)$ cost.
For the off-diagonal portions, there are three categories to consider:
(i) mapping amplitudes with fewer than four inactive indices amongst each other,
(ii) mapping amplitudes with four inactive indices amongst each other,
and (iii) mapping between these two sets of amplitudes.
For case (i), the amplitudes have at most three indices whose ranges are
$O(N)$, the rest having ranges that are $O(1)$.
As the zeroth order Hamiltonian is one-body, its off-diagonal part could
in principle contract with one of the $O(N)$ indices, leading to
terms that at first appear to have $O(N^4)$ scaling.
Examples of such terms include
\begin{align}
    \label{eqn:off-diag-virt}
    \sum_a t^{a \tilde{b}}_{ij} f_{ac} & \rightarrow t^{c \tilde{b}}_{ij} \\
    \label{eqn:off-diag-occ}
    \sum_j t^{a \tilde{b}}_{ij} f_{jk} & \rightarrow t^{a \tilde{b}}_{ik}
\end{align}
where $i$, $j$, and $k$ are inactive occupied orbitals, $a$ and $c$
are inactive virtuals, $\tilde{b}$ is an active virtual orbital,
and $f$ is the Fock matrix.
In practice, however, such terms have no worse than cubic scaling.
For these amplitudes, we use a canonical basis for the inactive virtual
orbitals, so $f_{ac} = \delta_{ac} f_{aa}$ in Eq.\ (\ref{eqn:off-diag-virt}).
The inactive occupied orbitals are localized instead of canonical,
and so although the Fock matrix between them is not diagonal,
there will asymptotically be only $O(1)$ values of $k$ for which
$f_{ik}$ is non-negligible for a given $i$, leading terms like the
one in Eq.\ (\ref{eqn:off-diag-occ}) to have cubic scaling as well.

Case (ii) is more straightforward, as these quadruply-inactive amplitudes
are the ones treated via DLPNO, and so there are only $O(N^2)$ such
amplitudes to begin with (note that our initial implementation provides
each inactive occupied pair with at least some PNOs).
Thus, their diagonal terms have quadratic scaling, and their virtual-to-virtual
off diagonal terms do as well as each inactive occupied pair's virtual space
is built from $O(1)$ PNOs.
As discussed above, the off-diagonal occupied-to-occupied block of the Fock
matrix will be sparse in the space of inactive occupieds because of their
locality, and so the occupied-to-occupied off-diagonal terms for case (ii)
will have quadratic scaling as well.

For case (iii) --- mappings between the amplitudes of cases (i) and (ii) ---
all terms also end up with cubic or better scaling.
For terms involving the virtual-to-virtual off-diagonal Fock elements,
the input and output amplitudes have the same two inactive occupied
indices.
The case (ii) amplitudes have two $O(1)$-range PNO virtuals,
while the case (i) amplitudes in such terms have at most one $O(N)$-range
virtual index (otherwise they would have too many inactive indices to
be case (i) amplitudes).
Thus, these virtual-to-virtual terms have at most three $O(N)$-range
indices in them, and so are no worse than cubic scaling.
For terms involving occupied-to-occupied off-diagonal Fock elements,
consider first Fock elements between inactive occupieds.
These produce nonzero case (iii) mappings, because the various PNOs
are not orthogonal to the other virtual orbitals, and so
amplitudes like $t^{uv}_{ij}$ and $t^{a\tilde{b}}_{ik}$
(where $u$ and $v$ label PNOs) are connected by $f_{jk}$.
Due to the aforementioned sparsity of this part of the Fock matrix,
terms mapping between these types of amplitudes have at most
three $O(N)$-range indices, and so scale cubically.

The final possibility is when a case (iii) term involves
Fock elements like $f_{j\tilde{k}}$ between inactive and active
occupieds, respectively.
These terms create nonzero mappings between amplitudes like
$t^{uv}_{ij}$ and $t^{ab}_{i\tilde{k}}$, which, at first glance,
appear to contain four $O(N)$-range indices.
However, although we have not explicitly localized the active
occupied orbitals, in the limit of a large system we expect the
orbitals that are strongly affected by the excitation to
naturally localize near the excitation's hole
or particle orbitals, and so, at least asymptotically, there
will only be $O(1)$ non-negligible values $f_{i\tilde{k}}$ in the
inactive-occupied-active-occupied block of the Fock matrix.
Thus, each term in this final group has at most three indices
with $O(N)$ ranges, completing the formal argument for why
the matrix vector multiplication needed by the conjugate gradient
solver will have cubic cost scaling.
Note that, although triples amplitudes were not used as explicit
examples in this argument, each block of ESMP2 triples differs from
a corresponding block of doubles by the addition of two $O(1)$-range
indices, and so their scaling analysis works out the same way.

The other task that we need to show cubic scaling for is the
evaluation of the linear equation's right hand side,
i.e.\ the elements of $\vec{r}$ from
Eq.\ (\ref{eqn:rhs-elems}).
This task boils down to evaluating various blocks of
electron repulsion integrals involving active and inactive
occupied and virtual indices.
In this study, we begin with a block-sparse \cite{rubensson2005systematic}
Cholesky decomposition \cite{koch2003reduced}
of the $(11|22)$-ordered atomic orbital integrals,
\begin{align}
    (xy|zw) = \sum_J (xy|J) (zw|J),
\end{align}
in which $x$, $y$, $z$ and $w$ label atomic orbitals
and $J$, whose range will be $O(N)$, indexes the Cholesky factors.
Due to the locality of atomic orbitals, only $O(N)$ $xy$
pairs contribute non-negligible integrals, and so the preparation
of the factors $(xy|J)$ will come at a cost that scales
no worse than cubically.
Although some applications of PNOs have achieved lower
scaling through local resolutions of the
identity,
\cite{guo2020linear}
we stick with a Cholesky factorization for now as it offers low enough
scaling that the evaluation of $\vec{r}$ is not our bottleneck
(in practice, the bottleneck is the conjugate gradient solve).
The Cholesky approach also provides tunable accuracy in
the decomposition via a single threshold \cite{koch2003reduced}
and avoids the need to choose an auxiliary basis.

This sparsity of $xy$ pairs ensures that, for a given value of $J$,
only $O(N)$ of the values $(xy|J)$ will be non-negligible.
Combined with the fact that each localized inactive occupied
orbital $i$ contains non-negligible contributions from only $O(1)$
atomic orbitals, transforming to the factor $(iy|J)$ has an
$O(N^2)$ cost and yields a tensor with $O(N^2)$ non-negligible elements.
In practice, we exploit this sparsity in a blocked fashion,
with the atomic orbitals for each 2nd row atom and its attendant
hydrogen atoms grouped into a block.
Thus, the arithmetic involved can be organized into
a collection of dense matrix multiplications,
ensuring that efficient linear algebra libraries
like BLAS can still be employed.
With the block-sparse factors $(iy|J)$ in hand, a sum over $J$
produces the tensor $(iy|jw)$ at $O(N^3)$ cost.
The $O(N^2)$ non-negligible elements in this tensor can then
be used to produce the integrals $(i\alpha|j\beta)$, where
$\alpha$ and $\beta$ label PNOs.
This last step has a total cost of $O(1)$ per $ij$ pair
due to the locality of the PNOs, and is carried out for each
of the $O(N^2)$ pairs.
With the $(i\alpha|j\beta)$ integrals in hand, the task of 
evaluating the elements of $\vec{r}$ that involve PNOs is complete.
In particular, note that, for the triples, the additional two TOP
indices are shared with the single excitation in the ESMF reference,
and so their $\vec{r}$ elements rely on the same $(i\alpha|j\beta)$
integrals as the doubles.

The elements of $\vec{r}$ that do not involve PNOs correspond to
amplitudes with at most three $O(N)$-range inactive indices.
The two blocks of integrals needed for these are
$(ia|j\tilde{b})$ and $(ia|\tilde{j}b)$, where $\tilde{j}$
and $\tilde{b}$ are active indices, $i$ and $j$ are (local)
inactive occupieds, and $a$ and $b$ are canonicalized
(so not local) inactive virtuals.
Exploiting block sparsity, the steps
\begin{align}
    (iy|J)(jw|J) & \rightarrow (iy|J) (j\tilde{b}|J)
                   \rightarrow (iy|j\tilde{b})
                   \rightarrow (ia|j\tilde{b}) 
\end{align}
have costs that scale as $O(N^2)$, $O(N^3)$, and $O(N^3)$, respectively,
thus delivering the $(ia|j\tilde{b})$ integrals at the desired cost.
Similarly, the steps
\begin{align}
    (iy|J)(zw|J) & \rightarrow (iy|J)(\tilde{j}w|J) \notag \\
                 & \rightarrow (iy|\tilde{j}w) \notag \\
                 & \rightarrow (ia|\tilde{j}w) \notag \\
                 & \rightarrow (ia|\tilde{j}b)
\end{align}
have costs that scale as
$O(N)$, $O(N^2)$, $O(N^2)$, and $O(N^3)$, respectively,
producing the $(ia|\tilde{j}b)$ integrals at cubic cost.
Note that the low cost of the first step is due to the
natural localization of the active orbitals near the excitation
and the fact that there are only $O(1)$ active orbitals.
Specifically, for each value of the index $\tilde{j}$,
there are $O(1)$ atomic orbitals $z$ with non-negligible contributions,
and for each $z$ there are $O(1)$ values of $w$ that matter,
leaving $J$ as the only index with an $O(N)$ range in the
conversion in this first step.
In summary, the integrals needed for the elements of $\vec{r}$
corresponding to amplitudes with three inactive indices
can be obtained at cubic cost.
Similar approaches can be used for those with fewer than three
inactive indices, which, as they have fewer $O(N)$-range indices,
are even less expensive to evaluate.
Thus, all the elements of $\vec{r}$ can be evaluated for a cost
that grows asymptotically as $O(N^3)$.

\subsection{Computational details}
\label{sec:comp-detail}

Excitation energy calculations employed the cc-pVDZ basis.
\cite{dunning1989gaussian}
EOM-CCSD, CC2, and CC3 calculations were performed with Psi4.
\cite{psi4}
Integrals for EMLC-ESMP2 were evaluated using Libint.
\cite{Libint2}
In the calculations below, three levels of tightness were
employed for the EMLC settings, as detailed in
Table \ref{tab:settings}.
Primary CSFs are those whose TOP singular values
are above t$_{\mathrm{primary}}$.
A CSF and its corresponding occupied orbital and virtual orbital
(its two half-filled orbitals) are flagged as active if
any of the following are true:
a) its TOP singular value is above t$_{\mathrm{active}}$,
b) the square norm of the ESMF occupied or virtual orbital after projection into the RHF occupied or virtual
   space is less than t$_{\mathrm{overlap}}$, or
c) its TOP singular value is greater than that of a CSF
   flagged as active by t$_{\mathrm{overlap}}$.
Active orbitals (which include the primary orbitals) are not localized,
and amplitudes involving them are not subject to the PNO approximation.
CSFs with singular values below t$_{\mathrm{explicit}}$
are truncated from the ESMF reference before the perturbation theory
is applied, with their ESMP2 correlation effects estimated as
described in Section \ref{sec:first-order-space}.
The tolerance for the Cholesky decomposition of the electron repulsion integrals is
given by t$_{\mathrm{cholesky}}$.
In our EMLC MP2 and ESMP2 equations, off-diagonal elements of the Fock matrix with absolute
values less than t$_{\mathrm{fock}}$ are neglected.
The PNOs used for each pair of inactive occupied orbitals are determined via the approach of
Riplinger and Neese \cite{DLPNO-CCSD} according to the thresholds
t$_{\mathrm{cutPNO}}$, t$_{\mathrm{cutMKN}}$, and t$_{\mathrm{cutMKN2}}$.
Note that t$_{\mathrm{cutMKN2}}$ is the threshold for including
projected atomic orbitals (PAOs) from neighboring atoms on which
the original atom's PAOs have significant population. \cite{DLPNO-CCSD}
We use a slightly modified definition
of population for domain selection and Pipek-Mezey
localization: \cite{pipek1989fast}
a molecular orbital's population on an atom is defined
as the square norm of its projection into the subspace
of that atom's valence atomic orbitals (defined for cc-pVDZ
as the 1s on hydrogen and the 1s, 2s, and 2p orbitals
on second row atoms).
Finally, in this study, each pair is provided with at least
n$_{\mathrm{minPNO}}$ PNOs, even if all the PNO
occupations for that pair are below t$_{\mathrm{cutPNO}}$.
While relaxing this constraint could
drop the number of treated pairs from
quadratic to linear in large systems, our
current implementation assumes that all
pairs are present.

\begin{table}[t]
\caption{Settings used in EMLC calculations.
Energetic quantities are in Hartrees.
See text for details.}
\label{tab:settings}
\centering
\begin{tabular}{l c c c}
\hline
\hline
\multicolumn{1}{c}{\rule[5.0mm]{0mm}{0mm}Setting}
 & loose & medium & tight \\
\hline
\rule[5.3mm]{0mm}{0mm}Level Shift             & 0.2       & 0.2       & 0.2 \\
\rule[4.5mm]{0mm}{0mm}t$_{\mathrm{primary}}$  & 0.05      & 0.05      & 0.01 \\
\rule[4.5mm]{0mm}{0mm}t$_{\mathrm{active}}$   & 10$^{-3}$ & 10$^{-3}$ & 10$^{-4}$ \\
\rule[4.5mm]{0mm}{0mm}t$_{\mathrm{overlap}}$  & 0.99      & 0.99      & 0.99 \\
\rule[4.5mm]{0mm}{0mm}t$_{\mathrm{explicit}}$ & 0.02      & 0.02      & 10$^{-3}$ \\
\rule[4.5mm]{0mm}{0mm}t$_{\mathrm{cholesky}}$ & 10$^{-4}$ & 10$^{-4}$ & 10$^{-6}$ \\
\rule[4.5mm]{0mm}{0mm}t$_{\mathrm{fock}}$     & 10$^{-2}$ & 10$^{-2}$ & 0 \\
\rule[4.5mm]{0mm}{0mm}t$_{\mathrm{cutPNO}}$  & 10$^{-4}$ & 10$^{-6}$ & 10$^{-10}$ \\
\rule[4.5mm]{0mm}{0mm}t$_{\mathrm{cutMKN}}$  & 10$^{-3}$ & 10$^{-3}$ & 10$^{-5}$ \\
\rule[4.5mm]{0mm}{0mm}t$_{\mathrm{cutMKN2}}$ & 0.05      & 0.05      & 10$^{-3}$ \\
\rule[4.5mm]{0mm}{0mm}n$_{\mathrm{minPNO}}$   & 2         & 2         & 2 \\
\hline
\hline
\end{tabular}
\end{table}

\section{Results}
\label{sec:results}

\subsection{Cost Scaling}
\label{sec:cost-scaling}

\begin{figure}[t]
    % trim=left bottom right top, clip
    \includegraphics[width=0.47\textwidth, trim=20 50 10 20, clip]{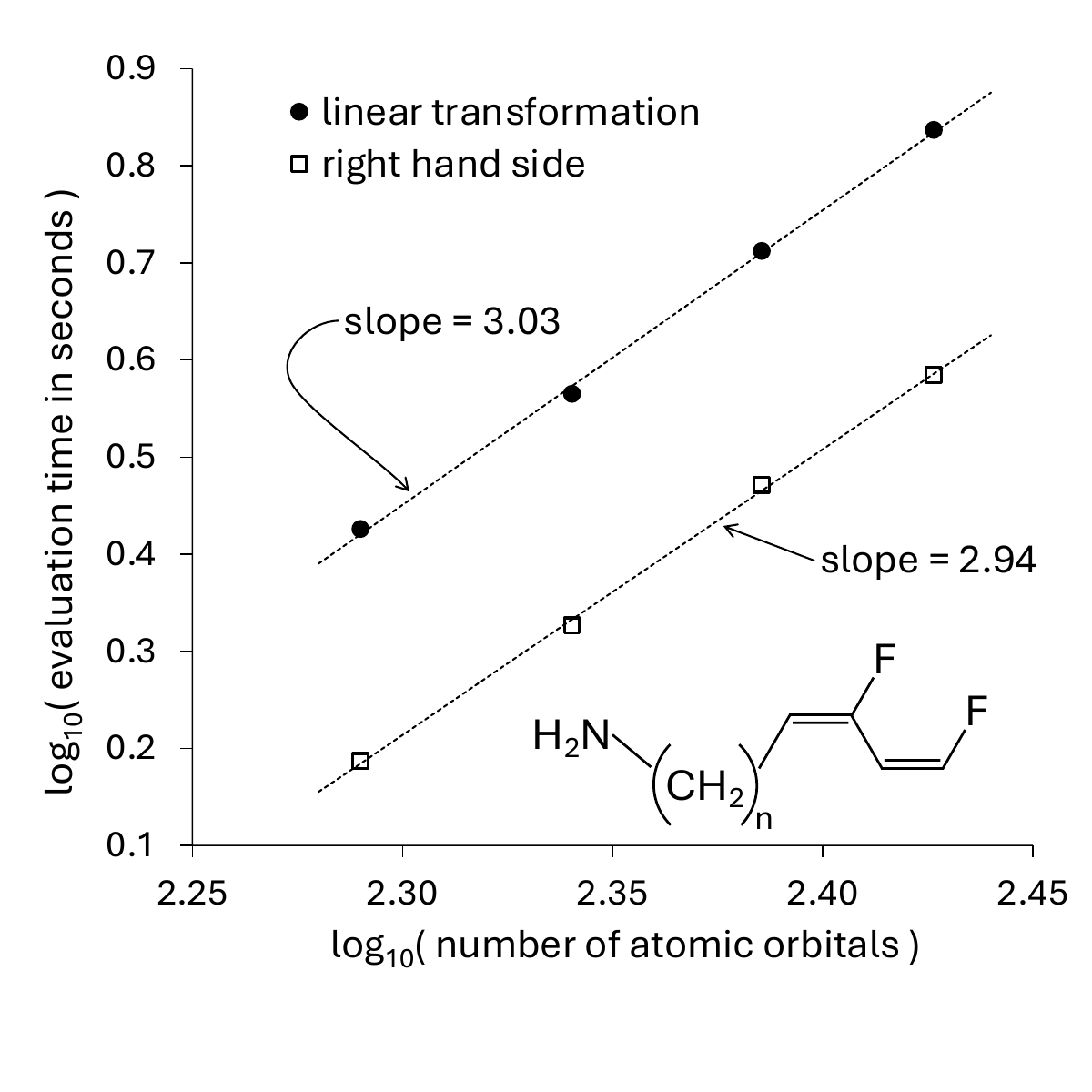}
    \caption{Cubic scaling demonstration for EMLC-ESMP2's
             linear transformation and right hand side in
             an $n\rightarrow\pi^{*}$ charge transfer state
             for n = 3, 4, 5, 6.
             Calculations used medium EMLC settings, the cc-pVDZ basis,
             and 56 cores across two Intel Xeon Gold 6330 processors.
             \label{fig:cost_scaling_n3}
             }
\end{figure}

As shown in Figure \ref{fig:cost_scaling_n3}, EMLC-ESMP2 achieves
its anticipated asymptotic cubic scaling in a system with as few
as ten non-hydrogen atoms.
In other words, there is a relatively early onset of the low scaling
regime.
This finding is especially encouraging given that
the method (with medium settings) is already significantly faster
(39 seconds vs 68 seconds on 56 Intel Xeon Gold 6330 cores)
than $N^6$-scaling EOM-CCSD in cc-pVDZ uracil, which has just
8 non-hydrogen atoms and 140 atomic orbitals.
Note that the low scaling is present in Figure \ref{fig:cost_scaling_n3}
despite the long range nature of the
excitation, in which an electron from the nitrogen lone pair is transferred
into an antibonding orbital in the fluorine-substituted $\pi$ system.
Even though the excitation is long range, only a small number of orbitals
are strongly affected, at least in the ways that matter for our EMLC approach
(remember, orbital relaxation effects have already been accounted for
in the ESMF reference).
Under the medium settings employed in this test,
the NH$_2$(CH$_2$)$_6$CHCFCHCHF molecule had only one CSF flagged
as primary in the TOP analysis of the ESMF state.
Just ten of its orbitals (five occupied and five virtual) were flagged as active,
leaving the remaining 46 occupied orbitals in the inactive category
that receives the local correlation treatment.
Of course, this early onset of low scaling is only useful if the method
offers reasonable accuracy, and so we will now turn our attention to two
tests: vertical excitation accuracy in simple valence excitations and
excitation and potential energy surface accuracy in a charge transfer example.

\subsection{Valence excitations}
\label{sec:valence-excitations}

Previous work \cite{clune2023thiel}
demonstrated that, with a level shift, the $N^5$-scaling version of
ESMP2 was able to achieve a 0.17 eV mean unsigned error (MUE)
for singlet single excitations on the Thiel set. \cite{thiel2008benchmark}
Although we do not attempt such an ambitious benchmarking effort in this study,
Table \ref{tab:test-set} shows results for a modest test set of
valence $n\hspace{0.7mm}$$\rightarrow$$\hspace{0.7mm}\pi^{*}$
and $\pi\hspace{0.7mm}$$\rightarrow$$\hspace{0.7mm}\pi^{*}$
singlet excitations.
Overall, we see that EMLC-ESMP2 achieves similar accuracy as was
seen on the Thiel set, with MUEs on the present test set of
0.15, 0.17, and 0.17 eV when using tight, medium, and loose settings, respectively.
Both CC2 and EOM-CCSD show better accuracy than EMLC-ESMP2 in the
current test set, which is a reversal compared to the methods' accuracy
ordering in the Thiel set tests. \cite{clune2023thiel}
The reason for this reversal is most likely in the choice of reference
data: this study used CC3/cc-pVDZ as the reference in all cases, whereas 
the Thiel set used a variety of methods and basis sets
informed by experimental spectra. \cite{thiel2008benchmark}
Nonetheless, the difference in overall accuracy on the present test set
between CC2, EOM-CCSD, and EMLC-ESMP2 is modest,
which is a significant achievement for a method with cubic scaling.

\begin{figure}[b!]
    % trim=left bottom right top, clip
    \includegraphics[width=0.47\textwidth, trim=10 20 15 26, clip]{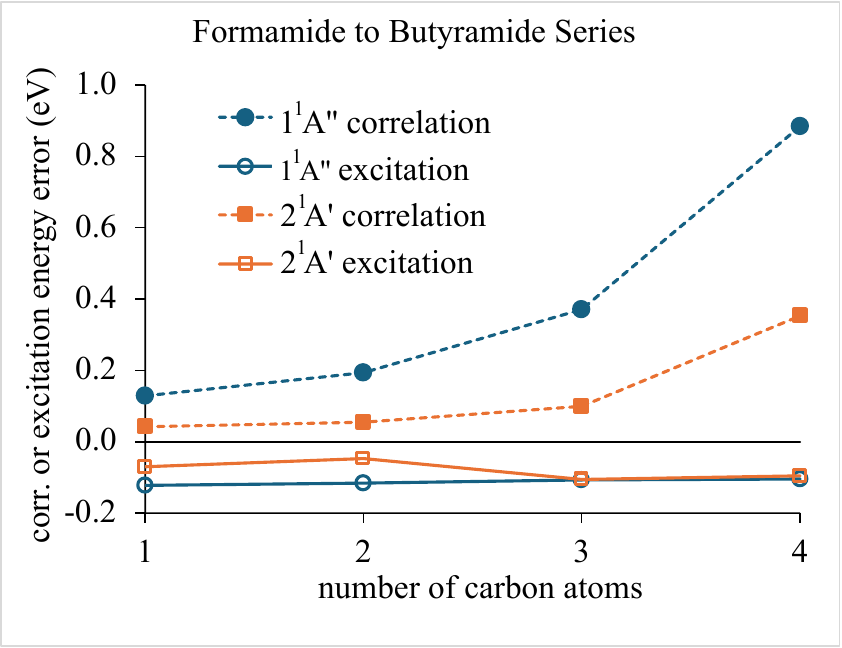}
    \caption{EMLC-ESMP2(loose) ground state correlation energy
             errors (vs MP2) and excitation energy
             errors (vs CC3) for the 1$^1$A$''$ and 2$^1$A$'$ states of
             formamide, acetamide, propanamide, and butyramide.
             The basis set is cc-pVDZ.
             \label{fig:amide-series}
             }
\end{figure}

\begin{table*}[t]
\caption{Excitation energies in eV.
         EMLC-ESMP2 results reported for tight, medium, and loose settings
         (see Table \ref{tab:settings}).
         Mean unsigned errors (MUE) are relative to CC3, and the
         \% correlation entry is for EMLC-MP2(loose).
}
\label{tab:test-set}
\centering
\begin{tabular}{l l l c c c c c c c}
\hline
\hline
\rule[5.3mm]{0mm}{0mm}Molecule &
State \hspace{3mm} &
Type \hspace{3.5mm} &
\hspace{2mm} CC3 \hspace{2mm} &
\hspace{1.4mm} CC2 \hspace{1.4mm} &
EOM-CCSD &
\hspace{1.4mm} tight \hspace{1.4mm} &
medium &
\hspace{1.4mm} loose \hspace{1.4mm} &
\% corr \\
\hline
\rule[5.3mm]{0mm}{0mm}formaldehyde      & 1$^1$A$_2$    & $n\hspace{0.7mm}$$\rightarrow$$\hspace{0.7mm}\pi^{*}$	  & 4.00   & 4.16   & 4.01     & 3.93    & 3.92     & 3.92    & 99.6    \\
\rule[4.5mm]{0mm}{0mm}acetone           & 1$^1$A$_2$    & $n\hspace{0.7mm}$$\rightarrow$$\hspace{0.7mm}\pi^{*}$	  & 4.43   & 4.56   & 4.45     & 4.35    & 4.35     & 4.35    & 99.3    \\
\rule[4.5mm]{0mm}{0mm}formaldehyde      & 1$^1$B$_1$    & $n\hspace{0.7mm}$$\rightarrow$$\hspace{0.7mm}\pi^{*}$	  & 9.29   & 9.47   & 9.36     & 9.19    & 9.15     & 9.15    & 99.9    \\
\rule[4.5mm]{0mm}{0mm}acetone           & 1$^1$B$_1$    & $n\hspace{0.7mm}$$\rightarrow$$\hspace{0.7mm}\pi^{*}$	  & 9.25   & 9.37   & 9.33     & 9.29    & 9.27     & 9.27    & 99.9    \\
\rule[4.5mm]{0mm}{0mm}formamide         & 1$^1$A$''$    & $n\hspace{0.7mm}$$\rightarrow$$\hspace{0.7mm}\pi^{*}$	  & 5.76   & 5.89   & 5.76     & 5.63    & 5.64     & 5.64    & 99.0    \\
\rule[4.5mm]{0mm}{0mm}acetamide         & 1$^1$A$''$    & $n\hspace{0.7mm}$$\rightarrow$$\hspace{0.7mm}\pi^{*}$	  & 5.79   & 5.87   & 5.80     & 5.67    & 5.67     & 5.68    & 98.8    \\
\rule[4.5mm]{0mm}{0mm}propanamide       & 1$^1$A$''$    & $n\hspace{0.7mm}$$\rightarrow$$\hspace{0.7mm}\pi^{*}$	  & 5.84   & 5.92   & 5.85     & 5.73    & 5.73     & 5.73    & 98.1    \\
\rule[4.5mm]{0mm}{0mm}butyramide        & 1$^1$A$''$    & $n\hspace{0.7mm}$$\rightarrow$$\hspace{0.7mm}\pi^{*}$	  & 5.87   & 5.94   & 5.87     & 5.76    & 5.76     & 5.76    & 96.3    \\
\rule[4.5mm]{0mm}{0mm}imidazole         & 1$^1$A$''$    & $n\hspace{0.7mm}$$\rightarrow$$\hspace{0.7mm}\pi^{*}$	  & 6.93   & 6.98   & 7.10     & 6.89    & 6.89     & 6.90    & 98.0    \\
\rule[4.5mm]{0mm}{0mm}uracil            & 1$^1$A$''$    & $n\hspace{0.7mm}$$\rightarrow$$\hspace{0.7mm}\pi^{*}$	  & 4.92   & 4.95   & 5.14     & 4.81    & 4.80     & 4.82    & 93.1    \\
\rule[4.5mm]{0mm}{0mm}formamide         & 2$^1$A$'$     & $\pi\hspace{0.7mm}$$\rightarrow$$\hspace{0.7mm}\pi^{*}$ & 7.58   & 7.65   & 7.85     & 7.54    & 7.51     & 7.51    & 99.7    \\
\rule[4.5mm]{0mm}{0mm}acetamide         & 2$^1$A$'$     & $\pi\hspace{0.7mm}$$\rightarrow$$\hspace{0.7mm}\pi^{*}$ & 7.57   & 7.44   & 7.87     & 7.58    & 7.52     & 7.52    & 99.7    \\
\rule[4.5mm]{0mm}{0mm}propanamide       & 2$^1$A$'$     & $\pi\hspace{0.7mm}$$\rightarrow$$\hspace{0.7mm}\pi^{*}$ & 7.59   & 7.49   & 7.87     & 7.53    & 7.48     & 7.49    & 99.5    \\
\rule[4.5mm]{0mm}{0mm}butyramide        & 2$^1$A$'$     & $\pi\hspace{0.7mm}$$\rightarrow$$\hspace{0.7mm}\pi^{*}$ & 7.60   & 7.50   & 7.89     & 7.54    & 7.50     & 7.50    & 98.5    \\
\rule[4.5mm]{0mm}{0mm}butadiene         & 1$^1$B$_u$    & $\pi\hspace{0.7mm}$$\rightarrow$$\hspace{0.7mm}\pi^{*}$ & 6.74   & 6.64   & 6.88     & 6.35    & 6.31     & 6.31    & 99.7    \\
\rule[4.5mm]{0mm}{0mm}hexatriene        & 1$^1$B$_u$    & $\pi\hspace{0.7mm}$$\rightarrow$$\hspace{0.7mm}\pi^{*}$ & 5.67   & 5.49   & 5.80     & 5.25    & 5.12     & 5.12    & 99.8    \\
\rule[4.5mm]{0mm}{0mm}cyclopentadiene   & 1$^1$B$_2$    & $\pi\hspace{0.7mm}$$\rightarrow$$\hspace{0.7mm}\pi^{*}$ & 5.81   & 5.77   & 5.94     & 5.45    & 5.40     & 5.40    & 99.8    \\
\rule[4.5mm]{0mm}{0mm}furan             & 1$^1$B$_2$    & $\pi\hspace{0.7mm}$$\rightarrow$$\hspace{0.7mm}\pi^{*}$ & 6.79   & 6.95   & 7.00     & 6.50    & 6.46     & 6.46    & 99.8    \\
\rule[4.5mm]{0mm}{0mm}pyridine          & 2$^1$A$_1$    & $\pi\hspace{0.7mm}$$\rightarrow$$\hspace{0.7mm}\pi^{*}$ & 6.94   & 6.97   & 7.02     & 7.27    & 7.19     & 7.19    & 99.7    \\
\hline
\multicolumn{3}{l}{\rule[5.3mm]{0mm}{0mm}MUE}                                                                     &        & 0.10   & 0.13     & 0.15    & 0.17     & 0.17    &         \\
\multicolumn{3}{l}{\rule[4.5mm]{0mm}{0mm}MUE, $n\hspace{0.7mm}$$\rightarrow$$\hspace{0.7mm}\pi^{*}$}              &        & 0.10   & 0.06     & 0.09    & 0.09     & 0.09    &         \\
\multicolumn{3}{l}{\rule[4.5mm]{0mm}{0mm}MUE, $\pi\hspace{0.7mm}$$\rightarrow$$\hspace{0.7mm}\pi^{*}$}            &        & 0.10   & 0.20     & 0.22    & 0.26     & 0.25    &         \\
\hline
\hline
\end{tabular}
\end{table*}

Digging into the details of the data in Table \ref{tab:test-set}, a number of
interesting observations can be made.
First, like EOM-CCSD, EMLC-ESMP2 appears to be more accurate for 
$n\hspace{0.7mm}$$\rightarrow$$\hspace{0.7mm}\pi^{*}$ excitations than it is for
$\pi\hspace{0.7mm}$$\rightarrow$$\hspace{0.7mm}\pi^{*}$ excitations.
Second, the states in which EMLC-ESMP2 displays the largest deviations from CC3
--- specifically, butadiene, hexatriene, cyclopentadiene, and furan ---
are all states for which the parent theory ($N^5$-scaling ESMP2)
displayed large errors in the Thiel set benchmark. \cite{clune2023thiel}
Thus, these errors do not appear to be caused by the use of the EMLC approximation.
Instead, they appear to be inherent to ESMP2 theory itself.
Finally, the degree to which the EMLC approach recovers the full MP2 or ESMP2
correlation energy does not appear to be related to the accuracy of the
excitation energy prediction.
Some states with high \% correlation recovery (hexatriene, cyclopentadiene, furan)
show relatively poor accuracy due to the limiations of ESMP2 itself,
while other states with high \% correlation recovery
(acetone 1$^1$B$_1$, acetamide 2$^1$A$'$)
show relatively good accuracy.
Likewise, some states with relatively low \% correlation recovery
(uracil, butyramide 2$^1$A$'$) display high accuracy.
Uracil in particular is an interesting example where, despite the modest
size of the molecule, the EMLC approach is already finding a large amount of
correlation that can be safely skipped while still maintaining
an accurate excitation energy.
Using loose settings, the ground state EMLC-MP2 calculation in uracil
misses 2.13 eV of correlation energy relative to canonical MP2, but the
excitation energy is only 0.1 eV less than the CC3 value.
Thus, even in these modest systems, we are already beginning to see EMLC's
intended cancellation of error play out successfully.

To further highlight this effect, Figure \ref{fig:amide-series} compares
the correlation energy error in the ground state to the excitation energy
accuracy for the states in the amide series.
As intended, as the alkane chain that is not involved in the excitation
gets longer, more and more of its correlation is simply skipped by the
EMLC approach, in which the inactive orbital pairs' local correlation
treatment is performed with an intentionally loose PNO threshold.
Again, the idea is that the correlation details for pairs uninvolved
in the excitation is likely to cancel out in the excitation energy,
and so it is only amplitudes involving active orbitals
that need a full correlation treatment.
This idea appears to be born out in the amide series, where we see
that the accuracy of the two states' excitation energies is essentially
unchanged across the series, despite the steady growth in the amount
of correlation energy that the EMLC approach (intentionally) neglects.

\begin{figure}[t]
    % trim=left bottom right top, clip
    \includegraphics[width=0.45\textwidth, trim=15 170 365 100, clip]{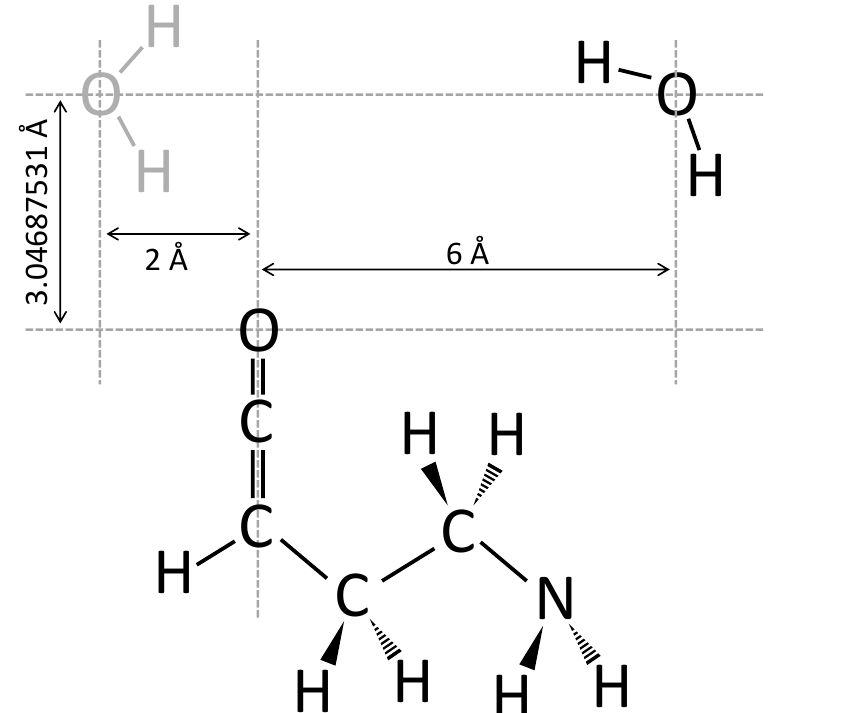}
    \caption{Initial (gray) and final (black) positions of the
      water's oxygen during the water flyby tests.
      Intersections between gray dashed lines are right angles.
      For each new geometry, the water's oxygen was moved 1 \AA\ rightwards,
      and then the water's hydrogen positions were determined by
      a C$_{\mathrm{S}}$ symmetric MP2/6-31G optimization with
      all other atoms held fixed.
      \label{fig:flyby-schematic}
    }
%    \\
\end{figure}

\begin{figure}[t]
    \captionsetup[subfigure]{labelformat=empty}
    \centering
    \vspace{7mm}
    \subfloat[]{
      % trim=left bottom right top, clip
      \includegraphics[width=0.45\textwidth, trim=20 5 10 30, clip]
                      {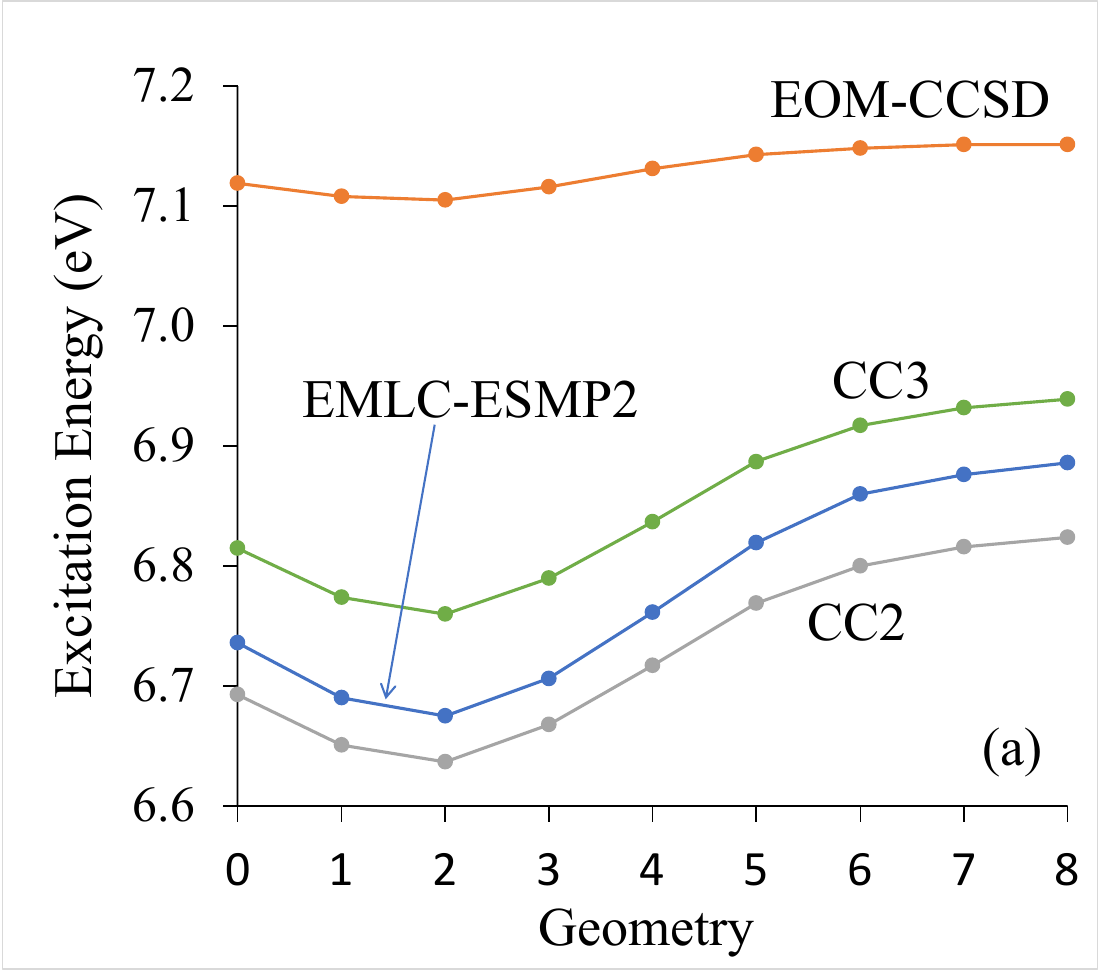}
    }
    \hfill
    \subfloat[]{
      % trim=left bottom right top, clip
      \includegraphics[width=0.45\textwidth, trim=20 5 10 30, clip]
                      {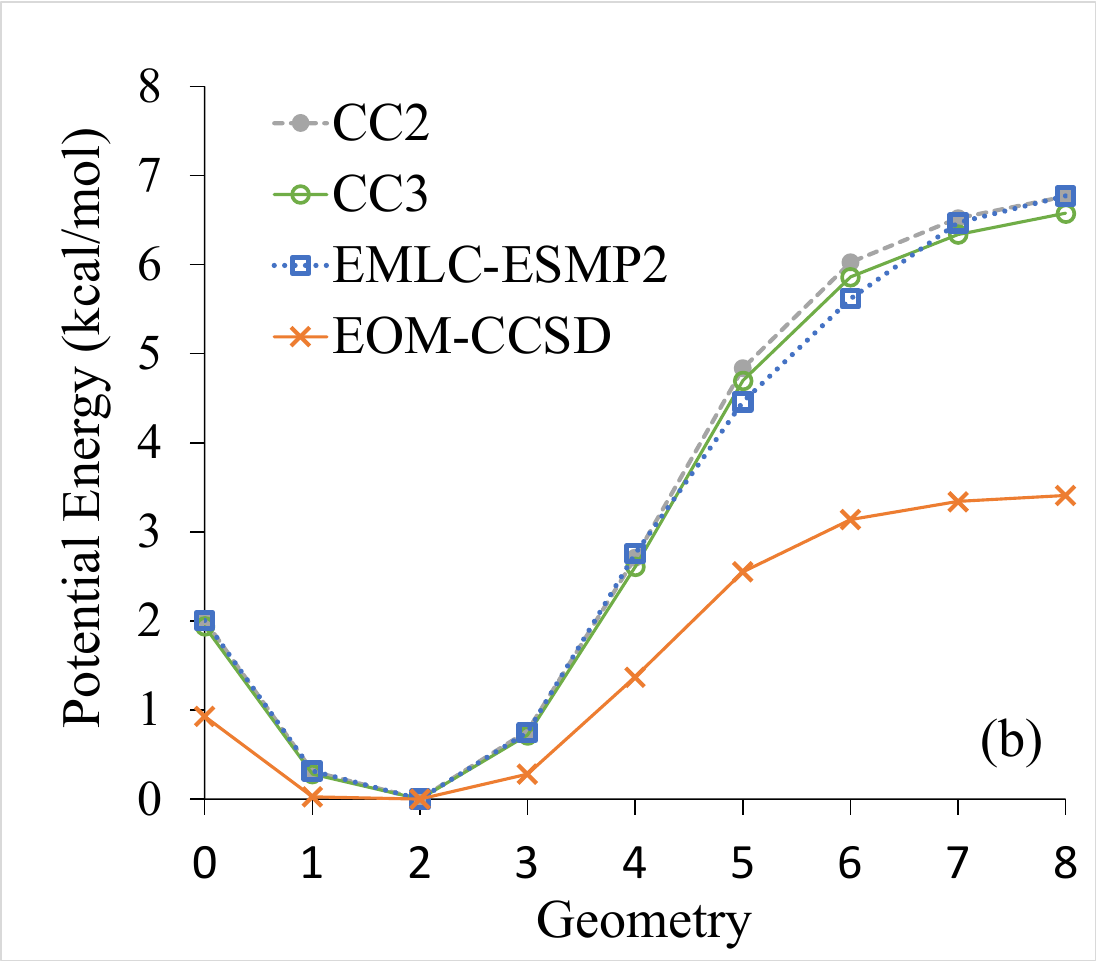}
    }
    \caption{(a) Excitation energies for the ${}^1$A$'$ nitrogen-to-CCO
             $n\hspace{0.7mm}$$\rightarrow\hspace{0.7mm}$$\pi^{*}$
             charge transfer excitation at the
             water flyby geometries from Figure \ref{fig:flyby-schematic}.
             (b) The corresponding excited state potential energy surfaces,
             each of which has been shifted to place its zero of energy
             at geometry 2.
             In both (a) and (b), the basis is cc-pVDZ and medium
             EMLC settings were used.
             \label{fig:gap-vs-geom}
             }
\end{figure}

%\FloatBarrier

\subsection{Water flyby test}
\label{sec:flyby}

We now turn our attention to a simple test that explores
how well different methods capture the effects that an explicit
solvent molecule has on a charge transfer excitation.
To do so, we drag a water molecule past the terminal oxygen
of the donor-bridge-acceptor
molecule shown in Figure \ref{fig:flyby-schematic}.
At each geometry (coordinates are in the SI),
we model the totally symmetric singlet charge transfer
excitation in which an electron is promoted from the nitrogen
lone pair into the CCO moiety's in-plane $\pi^{*}$ orbital.
In Figure \ref{fig:gap-vs-geom}, we see that EMLC-ESMP2
predicts the excitation energies within about 0.1 eV
and the excited state potential energy surface
within 0.5 kcal/mol of the CC3 reference.
This accuracy comes at a remarkably low cost:
each point took about 20 seconds with EMLC-ESMP2,
compared to two minutes with CC2, three minutes with EOM-CCSD,
and multiple hours with CC3.
Given the early onset of EMLC-ESMP2's cubic scaling observed
in the cost scaling analysis above, the
timings seen in this small water flyby
test underlines the cost advantage
that the EMLC approach offers.
When one considers TD-DFT's ongoing struggles \cite{mester2022charge}
with charge transfer excitations,
this preliminary data on the efficiency and accuracy
of EMLC-ESMP2 in a simple test of solvent-modulated
charge transfer is especially promising.

Looking into the details of the results, the relatively
poor performance of EOM-CCSD can be traced to its tendency to
overestimate charge transfer excitation energies due to its
limited treatment of orbital relaxation effects.
In this example, the overestimate leads to a prediction
that the $n\hspace{0.7mm}$$\rightarrow\hspace{0.7mm}$$\pi^{*}$
transition is sufficiently close in energy to a higher-lying,
non-charge-transfer transition so that the two mix in roughly
equal proportions.
The other methods, including CC3, do not predict such mixing,
presumably because they place the charge transfer transition
lower in energy.
In underestimating the degree of charge transfer character
in the excitation, EOM-CCSD underestimates the amount of charge
being built up on the CCO moiety and therefore the strength
of the water's interaction with it.
As a result, it substantially underestimates the rise in the
excited state PES as the water is moved away from that charge.

\section{Conclusions}

This study has explored an excitation matched local correlation
approach to excited-state-specific perturbation theory that seeks
to intentionally avoid evaluating portions of the correlation
energy expected to be the same in the ground and excited state.
We found that doing so yields a method that achieves cubic scaling
in remarkably small system sizes, even in the context of long
range charge transfer, and whose cost prefactor is low enough
to already be cost competitive with CC2 in the uracil molecule.
Tests on $n\hspace{0.7mm}$$\rightarrow$$\hspace{0.7mm}\pi^{*}$ and
$\pi\hspace{0.7mm}$$\rightarrow$$\hspace{0.7mm}\pi^{*}$
valence excitations, as well as on a charge transfer example
involving an explicit solvent molecule, reveal that the
approach's accuracy for excitation energies is essentially the
same as what was seen in previous benchmarking of the
parent ESMP2 method.
The new EMLC-ESMP2 appears to be particularly accurate
for $n\hspace{0.7mm}$$\rightarrow$$\hspace{0.7mm}\pi^{*}$
excitation energies, producing errors relative to CC3 of
about 0.1 eV in both valence and charge transfer examples.

In future, it will be very interesting to apply the EMLC
approach to Aufbau suppressed coupled cluster theory and its
corresponding second order perturbation theory.
That perturbation theory is considerably simpler than ESMP2
in terms of the number of tensor contractions needed to
evaluate its matrix-vector product, which should allow
for an EMLC treatment to achieve a much lower computational
prefactor than is possible for ESMP2 while still achieving
an early crossover to cubic scaling.
Applying the EMLC idea to the coupled cluster theory itself
would of course be more involved, but the history of local
correlation treatments in the ground state suggest that there
should be no fundamental barrier to doing so.
The payoff could be considerable, as it would combine
the high accuracy and systematic improvability of
Aufbau suppressed coupled cluster with the EMLC approach's
low cost and low scaling.

\section{Acknowledgments}

This work was supported by the National Science Foundation,
Award Number 2320936.
Calculations were performed using the Savio computational cluster resource provided by the Berkeley Research Computing program at the University of California, Berkeley.


%\bibliography{main}

\begin{thebibliography}{67}%
\makeatletter
\providecommand \@ifxundefined [1]{%
 \@ifx{#1\undefined}
}%
\providecommand \@ifnum [1]{%
 \ifnum #1\expandafter \@firstoftwo
 \else \expandafter \@secondoftwo
 \fi
}%
\providecommand \@ifx [1]{%
 \ifx #1\expandafter \@firstoftwo
 \else \expandafter \@secondoftwo
 \fi
}%
\providecommand \natexlab [1]{#1}%
\providecommand \enquote  [1]{``#1''}%
\providecommand \bibnamefont  [1]{#1}%
\providecommand \bibfnamefont [1]{#1}%
\providecommand \citenamefont [1]{#1}%
\providecommand \href@noop [0]{\@secondoftwo}%
\providecommand \href [0]{\begingroup \@sanitize@url \@href}%
\providecommand \@href[1]{\@@startlink{#1}\@@href}%
\providecommand \@@href[1]{\endgroup#1\@@endlink}%
\providecommand \@sanitize@url [0]{\catcode `\\12\catcode `\$12\catcode
  `\&12\catcode `\#12\catcode `\^12\catcode `\_12\catcode `\%12\relax}%
\providecommand \@@startlink[1]{}%
\providecommand \@@endlink[0]{}%
\providecommand \url  [0]{\begingroup\@sanitize@url \@url }%
\providecommand \@url [1]{\endgroup\@href {#1}{\urlprefix }}%
\providecommand \urlprefix  [0]{URL }%
\providecommand \Eprint [0]{\href }%
\providecommand \doibase [0]{http://dx.doi.org/}%
\providecommand \selectlanguage [0]{\@gobble}%
\providecommand \bibinfo  [0]{\@secondoftwo}%
\providecommand \bibfield  [0]{\@secondoftwo}%
\providecommand \translation [1]{[#1]}%
\providecommand \BibitemOpen [0]{}%
\providecommand \bibitemStop [0]{}%
\providecommand \bibitemNoStop [0]{.\EOS\space}%
\providecommand \EOS [0]{\spacefactor3000\relax}%
\providecommand \BibitemShut  [1]{\csname bibitem#1\endcsname}%
\let\auto@bib@innerbib\@empty
%</preamble>
\bibitem [{\citenamefont {Zheng}\ and\ \citenamefont
  {Cheng}(2019)}]{zheng2019performance}%
  \BibitemOpen
  \bibfield  {author} {\bibinfo {author} {\bibfnamefont {X.}~\bibnamefont
  {Zheng}}\ and\ \bibinfo {author} {\bibfnamefont {L.}~\bibnamefont {Cheng}},\
  }\bibfield  {title} {\enquote {\bibinfo {title} {Performance of
  delta-coupled-cluster methods for calculations of core-ionization energies of
  first-row elements},}\ }\href@noop {} {\bibfield  {journal} {\bibinfo
  {journal} {Journal of chemical theory and computation}\ }\textbf {\bibinfo
  {volume} {15}},\ \bibinfo {pages} {4945--4955} (\bibinfo {year}
  {2019})}\BibitemShut {NoStop}%
\bibitem [{\citenamefont {Lee}, \citenamefont {Small},\ and\ \citenamefont
  {Head-Gordon}(2019)}]{lee2019excited}%
  \BibitemOpen
  \bibfield  {author} {\bibinfo {author} {\bibfnamefont {J.}~\bibnamefont
  {Lee}}, \bibinfo {author} {\bibfnamefont {D.~W.}\ \bibnamefont {Small}}, \
  and\ \bibinfo {author} {\bibfnamefont {M.}~\bibnamefont {Head-Gordon}},\
  }\bibfield  {title} {\enquote {\bibinfo {title} {Excited states via coupled
  cluster theory without equation-of-motion methods: Seeking higher roots with
  application to doubly excited states and double core hole states},}\
  }\href@noop {} {\bibfield  {journal} {\bibinfo  {journal} {The Journal of
  chemical physics}\ }\textbf {\bibinfo {volume} {151}},\ \bibinfo {pages}
  {214103} (\bibinfo {year} {2019})}\BibitemShut {NoStop}%
\bibitem [{\citenamefont {Clune}, \citenamefont {Shea},\ and\ \citenamefont
  {Neuscamman}(2020)}]{N5-ESMP2}%
  \BibitemOpen
  \bibfield  {author} {\bibinfo {author} {\bibfnamefont {R.}~\bibnamefont
  {Clune}}, \bibinfo {author} {\bibfnamefont {J.~A.~R.}\ \bibnamefont {Shea}},
  \ and\ \bibinfo {author} {\bibfnamefont {E.}~\bibnamefont {Neuscamman}},\
  }\bibfield  {title} {\enquote {\bibinfo {title} {N-5-scaling
  excited-state-specific perturbation theory},}\ }\href {\doibase
  10.1021/acs.jctc.0c00308} {\bibfield  {journal} {\bibinfo  {journal} {J.
  Chem. Theory Comput.}\ }\textbf {\bibinfo {volume} {16}},\ \bibinfo {pages}
  {6132--6141} (\bibinfo {year} {2020})}\BibitemShut {NoStop}%
\bibitem [{\citenamefont {Tuckman}\ and\ \citenamefont
  {Neuscamman}(2024)}]{tuckman2024aufbau}%
  \BibitemOpen
  \bibfield  {author} {\bibinfo {author} {\bibfnamefont {H.}~\bibnamefont
  {Tuckman}}\ and\ \bibinfo {author} {\bibfnamefont {E.}~\bibnamefont
  {Neuscamman}},\ }\bibfield  {title} {\enquote {\bibinfo {title} {Aufbau
  suppressed coupled cluster theory for electronically excited states},}\
  }\href@noop {} {\bibfield  {journal} {\bibinfo  {journal} {Journal of
  Chemical Theory and Computation}\ }\textbf {\bibinfo {volume} {20}},\
  \bibinfo {pages} {2761--2773} (\bibinfo {year} {2024})}\BibitemShut {NoStop}%
\bibitem [{\citenamefont {Tuckman}, \citenamefont {Ma},\ and\ \citenamefont
  {Neuscamman}(2025)}]{tuckman2025improving}%
  \BibitemOpen
  \bibfield  {author} {\bibinfo {author} {\bibfnamefont {H.}~\bibnamefont
  {Tuckman}}, \bibinfo {author} {\bibfnamefont {Z.}~\bibnamefont {Ma}}, \ and\
  \bibinfo {author} {\bibfnamefont {E.}~\bibnamefont {Neuscamman}},\ }\bibfield
   {title} {\enquote {\bibinfo {title} {Improving aufbau suppressed coupled
  cluster through perturbative analysisclick to copy article link},}\
  }\href@noop {} {\bibfield  {journal} {\bibinfo  {journal} {Journal of
  Chemical Theory and Computation}\ }\textbf {\bibinfo {volume} {21}},\
  \bibinfo {pages} {3993} (\bibinfo {year} {2025})}\BibitemShut {NoStop}%
\bibitem [{\citenamefont {Pulay}(1983)}]{pulay1983localizability}%
  \BibitemOpen
  \bibfield  {author} {\bibinfo {author} {\bibfnamefont {P.}~\bibnamefont
  {Pulay}},\ }\bibfield  {title} {\enquote {\bibinfo {title} {Localizability of
  dynamic electron correlation},}\ }\href@noop {} {\bibfield  {journal}
  {\bibinfo  {journal} {Chemical physics letters}\ }\textbf {\bibinfo {volume}
  {100}},\ \bibinfo {pages} {151--154} (\bibinfo {year} {1983})}\BibitemShut
  {NoStop}%
\bibitem [{\citenamefont {Pulay}(1986)}]{pulay1986force}%
  \BibitemOpen
  \bibfield  {author} {\bibinfo {author} {\bibfnamefont {P.}~\bibnamefont
  {Pulay}},\ }\bibfield  {title} {\enquote {\bibinfo {title} {The force
  constants of benzene: Local many-body perturbation theory vs new
  experiment},}\ }\href@noop {} {\bibfield  {journal} {\bibinfo  {journal} {The
  Journal of chemical physics}\ }\textbf {\bibinfo {volume} {85}},\ \bibinfo
  {pages} {1703--1704} (\bibinfo {year} {1986})}\BibitemShut {NoStop}%
\bibitem [{\citenamefont {Pulay}\ and\ \citenamefont
  {Saeb{\o}}(1986)}]{pulay1986orbital}%
  \BibitemOpen
  \bibfield  {author} {\bibinfo {author} {\bibfnamefont {P.}~\bibnamefont
  {Pulay}}\ and\ \bibinfo {author} {\bibfnamefont {S.}~\bibnamefont
  {Saeb{\o}}},\ }\bibfield  {title} {\enquote {\bibinfo {title}
  {Orbital-invariant formulation and second-order gradient evaluation in
  m{\o}ller-plesset perturbation theory},}\ }\href@noop {} {\bibfield
  {journal} {\bibinfo  {journal} {Theoretica chimica acta}\ }\textbf {\bibinfo
  {volume} {69}},\ \bibinfo {pages} {357--368} (\bibinfo {year}
  {1986})}\BibitemShut {NoStop}%
\bibitem [{\citenamefont {Saeb{\o}}\ and\ \citenamefont
  {Pulay}(1987)}]{saebo1987fourth}%
  \BibitemOpen
  \bibfield  {author} {\bibinfo {author} {\bibfnamefont {S.}~\bibnamefont
  {Saeb{\o}}}\ and\ \bibinfo {author} {\bibfnamefont {P.}~\bibnamefont
  {Pulay}},\ }\bibfield  {title} {\enquote {\bibinfo {title} {Fourth-order
  m{\o}ller--plessett perturbation theory in the local correlation treatment.
  i. method},}\ }\href@noop {} {\bibfield  {journal} {\bibinfo  {journal} {The
  Journal of chemical physics}\ }\textbf {\bibinfo {volume} {86}},\ \bibinfo
  {pages} {914--922} (\bibinfo {year} {1987})}\BibitemShut {NoStop}%
\bibitem [{\citenamefont {Saeb{\o}}, \citenamefont {Tong},\ and\ \citenamefont
  {Pulay}(1993)}]{saebo1993efficient}%
  \BibitemOpen
  \bibfield  {author} {\bibinfo {author} {\bibfnamefont {S.}~\bibnamefont
  {Saeb{\o}}}, \bibinfo {author} {\bibfnamefont {W.}~\bibnamefont {Tong}}, \
  and\ \bibinfo {author} {\bibfnamefont {P.}~\bibnamefont {Pulay}},\ }\bibfield
   {title} {\enquote {\bibinfo {title} {Efficient elimination of basis set
  superposition errors by the local correlation method: Accurate ab initio
  studies of the water dimer},}\ }\href@noop {} {\bibfield  {journal} {\bibinfo
   {journal} {The Journal of chemical physics}\ }\textbf {\bibinfo {volume}
  {98}},\ \bibinfo {pages} {2170--2175} (\bibinfo {year} {1993})}\BibitemShut
  {NoStop}%
\bibitem [{\citenamefont {Hampel}\ and\ \citenamefont
  {Werner}(1996)}]{hampel1996local}%
  \BibitemOpen
  \bibfield  {author} {\bibinfo {author} {\bibfnamefont {C.}~\bibnamefont
  {Hampel}}\ and\ \bibinfo {author} {\bibfnamefont {H.-J.}\ \bibnamefont
  {Werner}},\ }\bibfield  {title} {\enquote {\bibinfo {title} {Local treatment
  of electron correlation in coupled cluster theory},}\ }\href@noop {}
  {\bibfield  {journal} {\bibinfo  {journal} {The Journal of chemical physics}\
  }\textbf {\bibinfo {volume} {104}},\ \bibinfo {pages} {6286--6297} (\bibinfo
  {year} {1996})}\BibitemShut {NoStop}%
\bibitem [{\citenamefont {Maslen}\ and\ \citenamefont
  {Head-Gordon}(1998)}]{maslen1998noniterative}%
  \BibitemOpen
  \bibfield  {author} {\bibinfo {author} {\bibfnamefont {P.~E.}\ \bibnamefont
  {Maslen}}\ and\ \bibinfo {author} {\bibfnamefont {M.}~\bibnamefont
  {Head-Gordon}},\ }\bibfield  {title} {\enquote {\bibinfo {title}
  {Noniterative local second order mo/ller--plesset theory: Convergence with
  local correlation space},}\ }\href@noop {} {\bibfield  {journal} {\bibinfo
  {journal} {The Journal of chemical physics}\ }\textbf {\bibinfo {volume}
  {109}},\ \bibinfo {pages} {7093--7099} (\bibinfo {year} {1998})}\BibitemShut
  {NoStop}%
\bibitem [{\citenamefont {Sch{\"u}tz}, \citenamefont {Hetzer},\ and\
  \citenamefont {Werner}(1999)}]{schutz1999low-I}%
  \BibitemOpen
  \bibfield  {author} {\bibinfo {author} {\bibfnamefont {M.}~\bibnamefont
  {Sch{\"u}tz}}, \bibinfo {author} {\bibfnamefont {G.}~\bibnamefont {Hetzer}},
  \ and\ \bibinfo {author} {\bibfnamefont {H.-J.}\ \bibnamefont {Werner}},\
  }\bibfield  {title} {\enquote {\bibinfo {title} {Low-order scaling local
  electron correlation methods. i. linear scaling local mp2},}\ }\href@noop {}
  {\bibfield  {journal} {\bibinfo  {journal} {The Journal of chemical physics}\
  }\textbf {\bibinfo {volume} {111}},\ \bibinfo {pages} {5691--5705} (\bibinfo
  {year} {1999})}\BibitemShut {NoStop}%
\bibitem [{\citenamefont {Sch{\"u}tz}(2000)}]{schutz2000low-III}%
  \BibitemOpen
  \bibfield  {author} {\bibinfo {author} {\bibfnamefont {M.}~\bibnamefont
  {Sch{\"u}tz}},\ }\bibfield  {title} {\enquote {\bibinfo {title} {Low-order
  scaling local electron correlation methods. iii. linear scaling local
  perturbative triples correction (t)},}\ }\href@noop {} {\bibfield  {journal}
  {\bibinfo  {journal} {The Journal of Chemical Physics}\ }\textbf {\bibinfo
  {volume} {113}},\ \bibinfo {pages} {9986--10001} (\bibinfo {year}
  {2000})}\BibitemShut {NoStop}%
\bibitem [{\citenamefont {Sch{\"u}tz}\ and\ \citenamefont
  {Werner}(2001)}]{schutz2001low-IV}%
  \BibitemOpen
  \bibfield  {author} {\bibinfo {author} {\bibfnamefont {M.}~\bibnamefont
  {Sch{\"u}tz}}\ and\ \bibinfo {author} {\bibfnamefont {H.-J.}\ \bibnamefont
  {Werner}},\ }\bibfield  {title} {\enquote {\bibinfo {title} {Low-order
  scaling local electron correlation methods. iv. linear scaling local
  coupled-cluster (lccsd)},}\ }\href@noop {} {\bibfield  {journal} {\bibinfo
  {journal} {The Journal of Chemical Physics}\ }\textbf {\bibinfo {volume}
  {114}},\ \bibinfo {pages} {661--681} (\bibinfo {year} {2001})}\BibitemShut
  {NoStop}%
\bibitem [{\citenamefont {Sch{\"u}tz}(2002)}]{schutz2002low-V}%
  \BibitemOpen
  \bibfield  {author} {\bibinfo {author} {\bibfnamefont {M.}~\bibnamefont
  {Sch{\"u}tz}},\ }\bibfield  {title} {\enquote {\bibinfo {title} {Low-order
  scaling local electron correlation methods. v. connected triples beyond (t):
  Linear scaling local ccsdt-1b},}\ }\href@noop {} {\bibfield  {journal}
  {\bibinfo  {journal} {The Journal of chemical physics}\ }\textbf {\bibinfo
  {volume} {116}},\ \bibinfo {pages} {8772--8785} (\bibinfo {year}
  {2002})}\BibitemShut {NoStop}%
\bibitem [{\citenamefont {Werner}, \citenamefont {Manby},\ and\ \citenamefont
  {Knowles}(2003)}]{werner2003fast}%
  \BibitemOpen
  \bibfield  {author} {\bibinfo {author} {\bibfnamefont {H.-J.}\ \bibnamefont
  {Werner}}, \bibinfo {author} {\bibfnamefont {F.~R.}\ \bibnamefont {Manby}}, \
  and\ \bibinfo {author} {\bibfnamefont {P.~J.}\ \bibnamefont {Knowles}},\
  }\bibfield  {title} {\enquote {\bibinfo {title} {Fast linear scaling
  second-order m{\o}ller-plesset perturbation theory (mp2) using local and
  density fitting approximations},}\ }\href@noop {} {\bibfield  {journal}
  {\bibinfo  {journal} {The Journal of chemical physics}\ }\textbf {\bibinfo
  {volume} {118}},\ \bibinfo {pages} {8149--8160} (\bibinfo {year}
  {2003})}\BibitemShut {NoStop}%
\bibitem [{\citenamefont {Kats}, \citenamefont {Korona},\ and\ \citenamefont
  {Schütz}(2006)}]{df_lcc2}%
  \BibitemOpen
  \bibfield  {author} {\bibinfo {author} {\bibfnamefont {D.}~\bibnamefont
  {Kats}}, \bibinfo {author} {\bibfnamefont {T.}~\bibnamefont {Korona}}, \ and\
  \bibinfo {author} {\bibfnamefont {M.}~\bibnamefont {Schütz}},\ }\bibfield
  {title} {\enquote {\bibinfo {title} {Local cc2 electronic excitation energies
  for large molecules with density fitting},}\ }\href {\doibase
  10.1063/1.2339021} {\bibfield  {journal} {\bibinfo  {journal} {J Chem Phys}\
  }\textbf {\bibinfo {volume} {125}},\ \bibinfo {pages} {104106} (\bibinfo
  {year} {2006})},\ \bibinfo {note} {kats, Danylo Korona, Tatiana Schütz,
  Martin 2006/9/27}\BibitemShut {NoStop}%
\bibitem [{\citenamefont {Kats}\ and\ \citenamefont
  {Schütz}(2009)}]{lt_df_lcc2}%
  \BibitemOpen
  \bibfield  {author} {\bibinfo {author} {\bibfnamefont {D.}~\bibnamefont
  {Kats}}\ and\ \bibinfo {author} {\bibfnamefont {M.}~\bibnamefont {Schütz}},\
  }\bibfield  {title} {\enquote {\bibinfo {title} {A multistate local coupled
  cluster cc2 response method based on the laplace transform},}\ }\href
  {\doibase 10.1063/1.3237134} {\bibfield  {journal} {\bibinfo  {journal} {J
  Chem Phys}\ }\textbf {\bibinfo {volume} {131}},\ \bibinfo {pages} {124117}
  (\bibinfo {year} {2009})},\ \bibinfo {note} {kats, Danylo Schütz, Martin
  2009/10/2}\BibitemShut {NoStop}%
\bibitem [{\citenamefont {Neese}, \citenamefont {Wennmohs},\ and\ \citenamefont
  {Hansen}(2009)}]{LPNO-CCSD}%
  \BibitemOpen
  \bibfield  {author} {\bibinfo {author} {\bibfnamefont {F.}~\bibnamefont
  {Neese}}, \bibinfo {author} {\bibfnamefont {F.}~\bibnamefont {Wennmohs}}, \
  and\ \bibinfo {author} {\bibfnamefont {A.}~\bibnamefont {Hansen}},\
  }\bibfield  {title} {\enquote {\bibinfo {title} {Efficient and accurate local
  approximations to coupled-electron pair approaches: An attempt to revive the
  pair natural orbital method},}\ }\href {\doibase 10.1063/1.3086717}
  {\bibfield  {journal} {\bibinfo  {journal} {J Chem Phys}\ }\textbf {\bibinfo
  {volume} {130}},\ \bibinfo {pages} {114108} (\bibinfo {year} {2009})},\
  \bibinfo {note} {neese, Frank Wennmohs, Frank Hansen, Andreas
  2009/3/26}\BibitemShut {NoStop}%
\bibitem [{\citenamefont {Riplinger}\ and\ \citenamefont
  {Neese}(2013)}]{DLPNO-CCSD}%
  \BibitemOpen
  \bibfield  {author} {\bibinfo {author} {\bibfnamefont {C.}~\bibnamefont
  {Riplinger}}\ and\ \bibinfo {author} {\bibfnamefont {F.}~\bibnamefont
  {Neese}},\ }\bibfield  {title} {\enquote {\bibinfo {title} {An efficient and
  near linear scaling pair natural orbital based local coupled cluster
  method},}\ }\href {\doibase 10.1063/1.4773581} {\bibfield  {journal}
  {\bibinfo  {journal} {J Chem Phys}\ }\textbf {\bibinfo {volume} {138}},\
  \bibinfo {pages} {034106} (\bibinfo {year} {2013})},\ \bibinfo {note}
  {riplinger, Christoph Neese, Frank 2013/1/25}\BibitemShut {NoStop}%
\bibitem [{\citenamefont {Neese}, \citenamefont {Hansen},\ and\ \citenamefont
  {Liakos}(2009)}]{neese2009efficient}%
  \BibitemOpen
  \bibfield  {author} {\bibinfo {author} {\bibfnamefont {F.}~\bibnamefont
  {Neese}}, \bibinfo {author} {\bibfnamefont {A.}~\bibnamefont {Hansen}}, \
  and\ \bibinfo {author} {\bibfnamefont {D.~G.}\ \bibnamefont {Liakos}},\
  }\bibfield  {title} {\enquote {\bibinfo {title} {Efficient and accurate
  approximations to the local coupled cluster singles doubles method using a
  truncated pair natural orbital basis},}\ }\href@noop {} {\bibfield  {journal}
  {\bibinfo  {journal} {The Journal of chemical physics}\ }\textbf {\bibinfo
  {volume} {131}},\ \bibinfo {pages} {064103} (\bibinfo {year}
  {2009})}\BibitemShut {NoStop}%
\bibitem [{\citenamefont {Auer}\ and\ \citenamefont
  {Nooijen}(2006)}]{auer2006dynamically}%
  \BibitemOpen
  \bibfield  {author} {\bibinfo {author} {\bibfnamefont {A.~A.}\ \bibnamefont
  {Auer}}\ and\ \bibinfo {author} {\bibfnamefont {M.}~\bibnamefont {Nooijen}},\
  }\bibfield  {title} {\enquote {\bibinfo {title} {Dynamically screened local
  correlation method using enveloping localized orbitals},}\ }\href@noop {}
  {\bibfield  {journal} {\bibinfo  {journal} {The Journal of chemical physics}\
  }\textbf {\bibinfo {volume} {125}},\ \bibinfo {pages} {024104} (\bibinfo
  {year} {2006})}\BibitemShut {NoStop}%
\bibitem [{\citenamefont {Russ}\ and\ \citenamefont
  {Crawford}(2008)}]{russ2008local}%
  \BibitemOpen
  \bibfield  {author} {\bibinfo {author} {\bibfnamefont {N.~J.}\ \bibnamefont
  {Russ}}\ and\ \bibinfo {author} {\bibfnamefont {T.~D.}\ \bibnamefont
  {Crawford}},\ }\bibfield  {title} {\enquote {\bibinfo {title} {Local
  correlation domains for coupled cluster theory: optical rotation and
  magnetic-field perturbations},}\ }\href@noop {} {\bibfield  {journal}
  {\bibinfo  {journal} {Physical Chemistry Chemical Physics}\ }\textbf
  {\bibinfo {volume} {10}},\ \bibinfo {pages} {3345--3352} (\bibinfo {year}
  {2008})}\BibitemShut {NoStop}%
\bibitem [{\citenamefont {Subotnik}\ and\ \citenamefont
  {Head-Gordon}(2005)}]{subotnik2005local}%
  \BibitemOpen
  \bibfield  {author} {\bibinfo {author} {\bibfnamefont {J.~E.}\ \bibnamefont
  {Subotnik}}\ and\ \bibinfo {author} {\bibfnamefont {M.}~\bibnamefont
  {Head-Gordon}},\ }\bibfield  {title} {\enquote {\bibinfo {title} {A local
  correlation model that yields intrinsically smooth potential-energy
  surfaces},}\ }\href@noop {} {\bibfield  {journal} {\bibinfo  {journal} {The
  Journal of chemical physics}\ }\textbf {\bibinfo {volume} {123}},\ \bibinfo
  {pages} {064108} (\bibinfo {year} {2005})}\BibitemShut {NoStop}%
\bibitem [{\citenamefont {Subotnik}\ and\ \citenamefont
  {Head-Gordon}(2008)}]{subotnik2008exploring}%
  \BibitemOpen
  \bibfield  {author} {\bibinfo {author} {\bibfnamefont {J.~E.}\ \bibnamefont
  {Subotnik}}\ and\ \bibinfo {author} {\bibfnamefont {M.}~\bibnamefont
  {Head-Gordon}},\ }\bibfield  {title} {\enquote {\bibinfo {title} {Exploring
  the accuracy of relative molecular energies with local correlation theory},}\
  }\href@noop {} {\bibfield  {journal} {\bibinfo  {journal} {Journal of
  Physics: Condensed Matter}\ }\textbf {\bibinfo {volume} {20}},\ \bibinfo
  {pages} {294211} (\bibinfo {year} {2008})}\BibitemShut {NoStop}%
\bibitem [{\citenamefont {Yang}\ \emph {et~al.}(2012)\citenamefont {Yang},
  \citenamefont {Chan}, \citenamefont {Manby}, \citenamefont {Sch{\"u}tz},\
  and\ \citenamefont {Werner}}]{yang2012orbital}%
  \BibitemOpen
  \bibfield  {author} {\bibinfo {author} {\bibfnamefont {J.}~\bibnamefont
  {Yang}}, \bibinfo {author} {\bibfnamefont {G.~K.}\ \bibnamefont {Chan}},
  \bibinfo {author} {\bibfnamefont {F.~R.}\ \bibnamefont {Manby}}, \bibinfo
  {author} {\bibfnamefont {M.}~\bibnamefont {Sch{\"u}tz}}, \ and\ \bibinfo
  {author} {\bibfnamefont {H.-J.}\ \bibnamefont {Werner}},\ }\bibfield  {title}
  {\enquote {\bibinfo {title} {The orbital-specific-virtual local coupled
  cluster singles and doubles method},}\ }\href@noop {} {\bibfield  {journal}
  {\bibinfo  {journal} {The Journal of Chemical Physics}\ }\textbf {\bibinfo
  {volume} {136}},\ \bibinfo {pages} {144105} (\bibinfo {year}
  {2012})}\BibitemShut {NoStop}%
\bibitem [{\citenamefont {Sparta}\ and\ \citenamefont
  {Neese}(2014)}]{sparta2014chemical}%
  \BibitemOpen
  \bibfield  {author} {\bibinfo {author} {\bibfnamefont {M.}~\bibnamefont
  {Sparta}}\ and\ \bibinfo {author} {\bibfnamefont {F.}~\bibnamefont {Neese}},\
  }\bibfield  {title} {\enquote {\bibinfo {title} {Chemical applications
  carried out by local pair natural orbital based coupled-cluster methods},}\
  }\href@noop {} {\bibfield  {journal} {\bibinfo  {journal} {Chemical Society
  Reviews}\ }\textbf {\bibinfo {volume} {43}},\ \bibinfo {pages} {5032--5041}
  (\bibinfo {year} {2014})}\BibitemShut {NoStop}%
\bibitem [{\citenamefont {Kats}(2014)}]{kats2014speeding}%
  \BibitemOpen
  \bibfield  {author} {\bibinfo {author} {\bibfnamefont {D.}~\bibnamefont
  {Kats}},\ }\bibfield  {title} {\enquote {\bibinfo {title} {Speeding up local
  correlation methods},}\ }\href@noop {} {\bibfield  {journal} {\bibinfo
  {journal} {The Journal of Chemical Physics}\ }\textbf {\bibinfo {volume}
  {141}},\ \bibinfo {pages} {244101} (\bibinfo {year} {2014})}\BibitemShut
  {NoStop}%
\bibitem [{\citenamefont {Kats}\ and\ \citenamefont
  {Werner}(2019)}]{pno_ms_caspt2}%
  \BibitemOpen
  \bibfield  {author} {\bibinfo {author} {\bibfnamefont {D.}~\bibnamefont
  {Kats}}\ and\ \bibinfo {author} {\bibfnamefont {H.~J.}\ \bibnamefont
  {Werner}},\ }\bibfield  {title} {\enquote {\bibinfo {title} {Multi-state
  local complete active space second-order perturbation theory using pair
  natural orbitals (pno-ms-caspt2)},}\ }\href {\doibase 10.1063/1.5097644}
  {\bibfield  {journal} {\bibinfo  {journal} {J Chem Phys}\ }\textbf {\bibinfo
  {volume} {150}},\ \bibinfo {pages} {214107} (\bibinfo {year} {2019})},\
  \bibinfo {note} {kats, Daniel Werner, Hans-Joachim 2019/6/10}\BibitemShut
  {NoStop}%
\bibitem [{\citenamefont {Liakos}\ \emph {et~al.}(2015)\citenamefont {Liakos},
  \citenamefont {Sparta}, \citenamefont {Kesharwani}, \citenamefont {Martin},\
  and\ \citenamefont {Neese}}]{liakos2015exploring}%
  \BibitemOpen
  \bibfield  {author} {\bibinfo {author} {\bibfnamefont {D.~G.}\ \bibnamefont
  {Liakos}}, \bibinfo {author} {\bibfnamefont {M.}~\bibnamefont {Sparta}},
  \bibinfo {author} {\bibfnamefont {M.~K.}\ \bibnamefont {Kesharwani}},
  \bibinfo {author} {\bibfnamefont {J.~M.}\ \bibnamefont {Martin}}, \ and\
  \bibinfo {author} {\bibfnamefont {F.}~\bibnamefont {Neese}},\ }\bibfield
  {title} {\enquote {\bibinfo {title} {Exploring the accuracy limits of local
  pair natural orbital coupled-cluster theory},}\ }\href@noop {} {\bibfield
  {journal} {\bibinfo  {journal} {Journal of chemical theory and computation}\
  }\textbf {\bibinfo {volume} {11}},\ \bibinfo {pages} {1525--1539} (\bibinfo
  {year} {2015})}\BibitemShut {NoStop}%
\bibitem [{\citenamefont {Riplinger}\ \emph {et~al.}(2016)\citenamefont
  {Riplinger}, \citenamefont {Pinski}, \citenamefont {Becker}, \citenamefont
  {Valeev},\ and\ \citenamefont {Neese}}]{riplinger2016sparse}%
  \BibitemOpen
  \bibfield  {author} {\bibinfo {author} {\bibfnamefont {C.}~\bibnamefont
  {Riplinger}}, \bibinfo {author} {\bibfnamefont {P.}~\bibnamefont {Pinski}},
  \bibinfo {author} {\bibfnamefont {U.}~\bibnamefont {Becker}}, \bibinfo
  {author} {\bibfnamefont {E.~F.}\ \bibnamefont {Valeev}}, \ and\ \bibinfo
  {author} {\bibfnamefont {F.}~\bibnamefont {Neese}},\ }\bibfield  {title}
  {\enquote {\bibinfo {title} {Sparse maps—a systematic infrastructure for
  reduced-scaling electronic structure methods. ii. linear scaling domain based
  pair natural orbital coupled cluster theory},}\ }\href@noop {} {\bibfield
  {journal} {\bibinfo  {journal} {The Journal of chemical physics}\ }\textbf
  {\bibinfo {volume} {144}},\ \bibinfo {pages} {024109} (\bibinfo {year}
  {2016})}\BibitemShut {NoStop}%
\bibitem [{\citenamefont {Tew}(2019)}]{tew2019principal}%
  \BibitemOpen
  \bibfield  {author} {\bibinfo {author} {\bibfnamefont {D.~P.}\ \bibnamefont
  {Tew}},\ }\bibfield  {title} {\enquote {\bibinfo {title} {Principal domains
  in local correlation theory},}\ }\href@noop {} {\bibfield  {journal}
  {\bibinfo  {journal} {Journal of Chemical Theory and Computation}\ }\textbf
  {\bibinfo {volume} {15}},\ \bibinfo {pages} {6597--6606} (\bibinfo {year}
  {2019})}\BibitemShut {NoStop}%
\bibitem [{\citenamefont {Guo}\ \emph {et~al.}(2020)\citenamefont {Guo},
  \citenamefont {Riplinger}, \citenamefont {Liakos}, \citenamefont {Becker},
  \citenamefont {Saitow},\ and\ \citenamefont {Neese}}]{guo2020linear}%
  \BibitemOpen
  \bibfield  {author} {\bibinfo {author} {\bibfnamefont {Y.}~\bibnamefont
  {Guo}}, \bibinfo {author} {\bibfnamefont {C.}~\bibnamefont {Riplinger}},
  \bibinfo {author} {\bibfnamefont {D.~G.}\ \bibnamefont {Liakos}}, \bibinfo
  {author} {\bibfnamefont {U.}~\bibnamefont {Becker}}, \bibinfo {author}
  {\bibfnamefont {M.}~\bibnamefont {Saitow}}, \ and\ \bibinfo {author}
  {\bibfnamefont {F.}~\bibnamefont {Neese}},\ }\bibfield  {title} {\enquote
  {\bibinfo {title} {Linear scaling perturbative triples correction
  approximations for open-shell domain-based local pair natural orbital coupled
  cluster singles and doubles theory [dlpno-ccsd (t/t)]},}\ }\href@noop {}
  {\bibfield  {journal} {\bibinfo  {journal} {The Journal of chemical physics}\
  }\textbf {\bibinfo {volume} {152}},\ \bibinfo {pages} {024116} (\bibinfo
  {year} {2020})}\BibitemShut {NoStop}%
\bibitem [{\citenamefont {Wang}\ \emph {et~al.}(2023)\citenamefont {Wang},
  \citenamefont {Aldossary}, \citenamefont {Shi}, \citenamefont {Liu},
  \citenamefont {Li},\ and\ \citenamefont {Head-Gordon}}]{wang2023local}%
  \BibitemOpen
  \bibfield  {author} {\bibinfo {author} {\bibfnamefont {Z.}~\bibnamefont
  {Wang}}, \bibinfo {author} {\bibfnamefont {A.}~\bibnamefont {Aldossary}},
  \bibinfo {author} {\bibfnamefont {T.}~\bibnamefont {Shi}}, \bibinfo {author}
  {\bibfnamefont {Y.}~\bibnamefont {Liu}}, \bibinfo {author} {\bibfnamefont
  {X.~S.}\ \bibnamefont {Li}}, \ and\ \bibinfo {author} {\bibfnamefont
  {M.}~\bibnamefont {Head-Gordon}},\ }\bibfield  {title} {\enquote {\bibinfo
  {title} {Local second-order m{\o}ller--plesset theory with a single threshold
  using orthogonal virtual orbitals: Theory, implementation, and assessment},}\
  }\href@noop {} {\bibfield  {journal} {\bibinfo  {journal} {Journal of
  Chemical Theory and Computation}\ }\textbf {\bibinfo {volume} {19}},\
  \bibinfo {pages} {7577--7591} (\bibinfo {year} {2023})}\BibitemShut {NoStop}%
\bibitem [{\citenamefont {Shi}\ \emph {et~al.}(2024)\citenamefont {Shi},
  \citenamefont {Wang}, \citenamefont {Aldossary}, \citenamefont {Liu},
  \citenamefont {Li},\ and\ \citenamefont {Head-Gordon}}]{shi2024local}%
  \BibitemOpen
  \bibfield  {author} {\bibinfo {author} {\bibfnamefont {T.}~\bibnamefont
  {Shi}}, \bibinfo {author} {\bibfnamefont {Z.}~\bibnamefont {Wang}}, \bibinfo
  {author} {\bibfnamefont {A.}~\bibnamefont {Aldossary}}, \bibinfo {author}
  {\bibfnamefont {Y.}~\bibnamefont {Liu}}, \bibinfo {author} {\bibfnamefont
  {X.~S.}\ \bibnamefont {Li}}, \ and\ \bibinfo {author} {\bibfnamefont
  {M.}~\bibnamefont {Head-Gordon}},\ }\bibfield  {title} {\enquote {\bibinfo
  {title} {Local second order m{\o}ller--plesset theory with a single threshold
  using orthogonal virtual orbitals: A distributed memory implementation},}\
  }\href@noop {} {\bibfield  {journal} {\bibinfo  {journal} {Journal of
  Chemical Theory and Computation}\ }\textbf {\bibinfo {volume} {20}},\
  \bibinfo {pages} {8010--8023} (\bibinfo {year} {2024})}\BibitemShut {NoStop}%
\bibitem [{\citenamefont {Nagy}(2024)}]{nagy2024state}%
  \BibitemOpen
  \bibfield  {author} {\bibinfo {author} {\bibfnamefont {P.~R.}\ \bibnamefont
  {Nagy}},\ }\bibfield  {title} {\enquote {\bibinfo {title} {State-of-the-art
  local correlation methods enable affordable gold standard quantum chemistry
  for up to hundreds of atoms},}\ }\href@noop {} {\bibfield  {journal}
  {\bibinfo  {journal} {Chemical Science}\ }\textbf {\bibinfo {volume} {15}},\
  \bibinfo {pages} {14556--14584} (\bibinfo {year} {2024})}\BibitemShut
  {NoStop}%
\bibitem [{\citenamefont {Yang}(2024)}]{yang2024making}%
  \BibitemOpen
  \bibfield  {author} {\bibinfo {author} {\bibfnamefont {J.}~\bibnamefont
  {Yang}},\ }\bibfield  {title} {\enquote {\bibinfo {title} {Making quantum
  chemistry compressive and expressive: Toward practical ab-initio
  simulation},}\ }\href@noop {} {\bibfield  {journal} {\bibinfo  {journal}
  {Wiley Interdisciplinary Reviews: Computational Molecular Science}\ }\textbf
  {\bibinfo {volume} {14}},\ \bibinfo {pages} {e1706} (\bibinfo {year}
  {2024})}\BibitemShut {NoStop}%
\bibitem [{\citenamefont {Russ}\ and\ \citenamefont
  {Crawford}(2004)}]{russ2004local}%
  \BibitemOpen
  \bibfield  {author} {\bibinfo {author} {\bibfnamefont {N.~J.}\ \bibnamefont
  {Russ}}\ and\ \bibinfo {author} {\bibfnamefont {T.~D.}\ \bibnamefont
  {Crawford}},\ }\bibfield  {title} {\enquote {\bibinfo {title} {Local
  correlation in coupled cluster calculations of molecular response
  properties},}\ }\href@noop {} {\bibfield  {journal} {\bibinfo  {journal}
  {Chemical physics letters}\ }\textbf {\bibinfo {volume} {400}},\ \bibinfo
  {pages} {104--111} (\bibinfo {year} {2004})}\BibitemShut {NoStop}%
\bibitem [{\citenamefont {Liang}\ \emph {et~al.}(2025)\citenamefont {Liang},
  \citenamefont {Zhang}, \citenamefont {Chan}, \citenamefont {Berkelbach},\
  and\ \citenamefont {Ye}}]{liang2025efficient}%
  \BibitemOpen
  \bibfield  {author} {\bibinfo {author} {\bibfnamefont {Y.~H.}\ \bibnamefont
  {Liang}}, \bibinfo {author} {\bibfnamefont {X.}~\bibnamefont {Zhang}},
  \bibinfo {author} {\bibfnamefont {G.~K.-L.}\ \bibnamefont {Chan}}, \bibinfo
  {author} {\bibfnamefont {T.~C.}\ \bibnamefont {Berkelbach}}, \ and\ \bibinfo
  {author} {\bibfnamefont {H.-Z.}\ \bibnamefont {Ye}},\ }\bibfield  {title}
  {\enquote {\bibinfo {title} {Efficient implementation of the random phase
  approximation with domain-based local pair natural orbitals},}\ }\href@noop
  {} {\bibfield  {journal} {\bibinfo  {journal} {Journal of Chemical Theory and
  Computation}\ }\textbf {\bibinfo {volume} {21}},\ \bibinfo {pages}
  {2918--2927} (\bibinfo {year} {2025})}\BibitemShut {NoStop}%
\bibitem [{\citenamefont {Baerends}, \citenamefont {Ellis},\ and\ \citenamefont
  {Ros}(1973)}]{baerends1973self}%
  \BibitemOpen
  \bibfield  {author} {\bibinfo {author} {\bibfnamefont {E.~J.}\ \bibnamefont
  {Baerends}}, \bibinfo {author} {\bibfnamefont {D.}~\bibnamefont {Ellis}}, \
  and\ \bibinfo {author} {\bibfnamefont {P.}~\bibnamefont {Ros}},\ }\bibfield
  {title} {\enquote {\bibinfo {title} {Self-consistent molecular
  hartree—fock—slater calculations i. the computational procedure},}\
  }\href@noop {} {\bibfield  {journal} {\bibinfo  {journal} {Chemical Physics}\
  }\textbf {\bibinfo {volume} {2}},\ \bibinfo {pages} {41--51} (\bibinfo {year}
  {1973})}\BibitemShut {NoStop}%
\bibitem [{\citenamefont {Nicklass}\ \emph {et~al.}(1995)\citenamefont
  {Nicklass}, \citenamefont {Dolg}, \citenamefont {Stoll},\ and\ \citenamefont
  {Preuss}}]{STU}%
  \BibitemOpen
  \bibfield  {author} {\bibinfo {author} {\bibfnamefont {A.}~\bibnamefont
  {Nicklass}}, \bibinfo {author} {\bibfnamefont {M.}~\bibnamefont {Dolg}},
  \bibinfo {author} {\bibfnamefont {H.}~\bibnamefont {Stoll}}, \ and\ \bibinfo
  {author} {\bibfnamefont {H.}~\bibnamefont {Preuss}},\ }\bibfield  {title}
  {\enquote {\bibinfo {title} {Ab initio energy‐adjusted pseudopotentials for
  the noble gases ne through xe: Calculation of atomic dipole and quadrupole
  polarizabilities},}\ }\href {\doibase 10.1063/1.468948} {\bibfield  {journal}
  {\bibinfo  {journal} {The Journal of Chemical Physics}\ }\textbf {\bibinfo
  {volume} {102}},\ \bibinfo {pages} {8942--8952} (\bibinfo {year}
  {1995})}\BibitemShut {NoStop}%
\bibitem [{\citenamefont {Burkatzki}, \citenamefont {Filippi},\ and\
  \citenamefont {Dolg}(2007)}]{BFD}%
  \BibitemOpen
  \bibfield  {author} {\bibinfo {author} {\bibfnamefont {M.}~\bibnamefont
  {Burkatzki}}, \bibinfo {author} {\bibfnamefont {C.}~\bibnamefont {Filippi}},
  \ and\ \bibinfo {author} {\bibfnamefont {M.}~\bibnamefont {Dolg}},\
  }\bibfield  {title} {\enquote {\bibinfo {title} {Energy-consistent
  pseudopotentials for quantum monte carlo calculations},}\ }\href {\doibase
  10.1063/1.2741534} {\bibfield  {journal} {\bibinfo  {journal} {The Journal of
  Chemical Physics}\ }\textbf {\bibinfo {volume} {126}},\ \bibinfo {pages}
  {234105} (\bibinfo {year} {2007})}\BibitemShut {NoStop}%
\bibitem [{\citenamefont {Trail}\ and\ \citenamefont {Needs}(2017)}]{cECP}%
  \BibitemOpen
  \bibfield  {author} {\bibinfo {author} {\bibfnamefont {J.~R.}\ \bibnamefont
  {Trail}}\ and\ \bibinfo {author} {\bibfnamefont {R.~J.}\ \bibnamefont
  {Needs}},\ }\bibfield  {title} {\enquote {\bibinfo {title} {Shape and energy
  consistent pseudopotentials for correlated electron systems},}\ }\href
  {\doibase 10.1063/1.4984046} {\bibfield  {journal} {\bibinfo  {journal} {The
  Journal of Chemical Physics}\ }\textbf {\bibinfo {volume} {146}},\ \bibinfo
  {pages} {204107} (\bibinfo {year} {2017})}\BibitemShut {NoStop}%
\bibitem [{\citenamefont {Wang}\ \emph {et~al.}(2022)\citenamefont {Wang},
  \citenamefont {Kincaid}, \citenamefont {Zhou}, \citenamefont {Annaberdiyev},
  \citenamefont {Bennett}, \citenamefont {Krogel},\ and\ \citenamefont
  {Mitas}}]{ccECP}%
  \BibitemOpen
  \bibfield  {author} {\bibinfo {author} {\bibfnamefont {G.}~\bibnamefont
  {Wang}}, \bibinfo {author} {\bibfnamefont {B.}~\bibnamefont {Kincaid}},
  \bibinfo {author} {\bibfnamefont {H.}~\bibnamefont {Zhou}}, \bibinfo {author}
  {\bibfnamefont {A.}~\bibnamefont {Annaberdiyev}}, \bibinfo {author}
  {\bibfnamefont {M.~C.}\ \bibnamefont {Bennett}}, \bibinfo {author}
  {\bibfnamefont {J.~T.}\ \bibnamefont {Krogel}}, \ and\ \bibinfo {author}
  {\bibfnamefont {L.}~\bibnamefont {Mitas}},\ }\bibfield  {title} {\enquote
  {\bibinfo {title} {A new generation of effective core potentials from
  correlated and spin–orbit calculations: Selected heavy elements},}\ }\href
  {\doibase 10.1063/5.0087300} {\bibfield  {journal} {\bibinfo  {journal} {The
  Journal of Chemical Physics}\ }\textbf {\bibinfo {volume} {157}},\ \bibinfo
  {pages} {054101} (\bibinfo {year} {2022})}\BibitemShut {NoStop}%
\bibitem [{\citenamefont {Miralles}\ \emph {et~al.}(1993)\citenamefont
  {Miralles}, \citenamefont {Castell}, \citenamefont {Caballol},\ and\
  \citenamefont {Malrieu}}]{miralles1993specific}%
  \BibitemOpen
  \bibfield  {author} {\bibinfo {author} {\bibfnamefont {J.}~\bibnamefont
  {Miralles}}, \bibinfo {author} {\bibfnamefont {O.}~\bibnamefont {Castell}},
  \bibinfo {author} {\bibfnamefont {R.}~\bibnamefont {Caballol}}, \ and\
  \bibinfo {author} {\bibfnamefont {J.-P.}\ \bibnamefont {Malrieu}},\
  }\bibfield  {title} {\enquote {\bibinfo {title} {Specific ci calculation of
  energy differences: Transition energies and bond energies},}\ }\href@noop {}
  {\bibfield  {journal} {\bibinfo  {journal} {Chemical physics}\ }\textbf
  {\bibinfo {volume} {172}},\ \bibinfo {pages} {33--43} (\bibinfo {year}
  {1993})}\BibitemShut {NoStop}%
\bibitem [{\citenamefont {Garc{\'\i}a}\ \emph {et~al.}(1995)\citenamefont
  {Garc{\'\i}a}, \citenamefont {Castell}, \citenamefont {Caballol},\ and\
  \citenamefont {Malrieu}}]{garcia1995iterative}%
  \BibitemOpen
  \bibfield  {author} {\bibinfo {author} {\bibfnamefont {V.}~\bibnamefont
  {Garc{\'\i}a}}, \bibinfo {author} {\bibfnamefont {O.}~\bibnamefont
  {Castell}}, \bibinfo {author} {\bibfnamefont {R.}~\bibnamefont {Caballol}}, \
  and\ \bibinfo {author} {\bibfnamefont {J.}~\bibnamefont {Malrieu}},\
  }\bibfield  {title} {\enquote {\bibinfo {title} {An iterative
  difference-dedicated configuration interaction. proposal and test studies},}\
  }\href@noop {} {\bibfield  {journal} {\bibinfo  {journal} {Chemical physics
  letters}\ }\textbf {\bibinfo {volume} {238}},\ \bibinfo {pages} {222--229}
  (\bibinfo {year} {1995})}\BibitemShut {NoStop}%
\bibitem [{\citenamefont {Chien}\ and\ \citenamefont
  {Zimmerman}(2017)}]{chien2017recovering}%
  \BibitemOpen
  \bibfield  {author} {\bibinfo {author} {\bibfnamefont {A.~D.}\ \bibnamefont
  {Chien}}\ and\ \bibinfo {author} {\bibfnamefont {P.~M.}\ \bibnamefont
  {Zimmerman}},\ }\bibfield  {title} {\enquote {\bibinfo {title} {Recovering
  dynamic correlation in spin flip configuration interaction through a
  difference dedicated approach},}\ }\href@noop {} {\bibfield  {journal}
  {\bibinfo  {journal} {The Journal of Chemical Physics}\ }\textbf {\bibinfo
  {volume} {146}} (\bibinfo {year} {2017})}\BibitemShut {NoStop}%
\bibitem [{\citenamefont {Gilbert}, \citenamefont {Besley},\ and\ \citenamefont
  {Gill}(2008)}]{gilbert2008self}%
  \BibitemOpen
  \bibfield  {author} {\bibinfo {author} {\bibfnamefont {A.~T.}\ \bibnamefont
  {Gilbert}}, \bibinfo {author} {\bibfnamefont {N.~A.}\ \bibnamefont {Besley}},
  \ and\ \bibinfo {author} {\bibfnamefont {P.~M.}\ \bibnamefont {Gill}},\
  }\bibfield  {title} {\enquote {\bibinfo {title} {Self-consistent field
  calculations of excited states using the maximum overlap method (mom)},}\
  }\href@noop {} {\bibfield  {journal} {\bibinfo  {journal} {The Journal of
  Physical Chemistry A}\ }\textbf {\bibinfo {volume} {112}},\ \bibinfo {pages}
  {13164--13171} (\bibinfo {year} {2008})}\BibitemShut {NoStop}%
\bibitem [{\citenamefont {Besley}, \citenamefont {Gilbert},\ and\ \citenamefont
  {Gill}(2009)}]{besley2009self}%
  \BibitemOpen
  \bibfield  {author} {\bibinfo {author} {\bibfnamefont {N.~A.}\ \bibnamefont
  {Besley}}, \bibinfo {author} {\bibfnamefont {A.~T.}\ \bibnamefont {Gilbert}},
  \ and\ \bibinfo {author} {\bibfnamefont {P.~M.}\ \bibnamefont {Gill}},\
  }\bibfield  {title} {\enquote {\bibinfo {title} {Self-consistent-field
  calculations of core excited states},}\ }\href@noop {} {\bibfield  {journal}
  {\bibinfo  {journal} {The Journal of chemical physics}\ }\textbf {\bibinfo
  {volume} {130}},\ \bibinfo {pages} {124308} (\bibinfo {year}
  {2009})}\BibitemShut {NoStop}%
\bibitem [{\citenamefont {Shea}\ and\ \citenamefont {Neuscamman}(2018)}]{ESMF}%
  \BibitemOpen
  \bibfield  {author} {\bibinfo {author} {\bibfnamefont {J.~A.~R.}\
  \bibnamefont {Shea}}\ and\ \bibinfo {author} {\bibfnamefont {E.}~\bibnamefont
  {Neuscamman}},\ }\bibfield  {title} {\enquote {\bibinfo {title}
  {Communication: A mean field platform for excited state quantum chemistry},}\
  }\href {\doibase 10.1063/1.5045056} {\bibfield  {journal} {\bibinfo
  {journal} {Journal of Chemical Physics}\ }\textbf {\bibinfo {volume} {149}}
  (\bibinfo {year} {2018}),\ 10.1063/1.5045056},\ \bibinfo {note} {shea,
  Jacqueline A. R. Neuscamman, Eric 1089-7690}\BibitemShut {NoStop}%
\bibitem [{\citenamefont {Shea}, \citenamefont {Gwin},\ and\ \citenamefont
  {Neuscamman}(2020)}]{GVP}%
  \BibitemOpen
  \bibfield  {author} {\bibinfo {author} {\bibfnamefont {J.~A.~R.}\
  \bibnamefont {Shea}}, \bibinfo {author} {\bibfnamefont {E.}~\bibnamefont
  {Gwin}}, \ and\ \bibinfo {author} {\bibfnamefont {E.}~\bibnamefont
  {Neuscamman}},\ }\bibfield  {title} {\enquote {\bibinfo {title} {A
  generalized variational principle with applications to excited state mean
  field theory},}\ }\href {\doibase 10.1021/acs.jctc.9b01105} {\bibfield
  {journal} {\bibinfo  {journal} {Journal of Chemical Theory and Computation}\
  }\textbf {\bibinfo {volume} {16}},\ \bibinfo {pages} {1526--1540} (\bibinfo
  {year} {2020})},\ \bibinfo {note} {shea, Jacqueline A. R. Gwin, Elise
  Neuscamman, Eric Shea, Jacqueline/G-1862-2015 Shea,
  Jacqueline/0000-0002-6179-1956 1549-9626}\BibitemShut {NoStop}%
\bibitem [{\citenamefont {Hardikar}\ and\ \citenamefont
  {Neuscamman}(2020)}]{hardikar2020self}%
  \BibitemOpen
  \bibfield  {author} {\bibinfo {author} {\bibfnamefont {T.~S.}\ \bibnamefont
  {Hardikar}}\ and\ \bibinfo {author} {\bibfnamefont {E.}~\bibnamefont
  {Neuscamman}},\ }\bibfield  {title} {\enquote {\bibinfo {title} {A
  self-consistent field formulation of excited state mean field theory},}\
  }\href@noop {} {\bibfield  {journal} {\bibinfo  {journal} {The Journal of
  chemical physics}\ }\textbf {\bibinfo {volume} {153}},\ \bibinfo {pages}
  {164108} (\bibinfo {year} {2020})}\BibitemShut {NoStop}%
\bibitem [{\citenamefont {Hait}\ and\ \citenamefont
  {Head-Gordon}(2020)}]{hait2020excited}%
  \BibitemOpen
  \bibfield  {author} {\bibinfo {author} {\bibfnamefont {D.}~\bibnamefont
  {Hait}}\ and\ \bibinfo {author} {\bibfnamefont {M.}~\bibnamefont
  {Head-Gordon}},\ }\bibfield  {title} {\enquote {\bibinfo {title} {Excited
  state orbital optimization via minimizing the square of the gradient: General
  approach and application to singly and doubly excited states via density
  functional theory},}\ }\href@noop {} {\bibfield  {journal} {\bibinfo
  {journal} {Journal of chemical theory and computation}\ }\textbf {\bibinfo
  {volume} {16}},\ \bibinfo {pages} {1699--1710} (\bibinfo {year}
  {2020})}\BibitemShut {NoStop}%
\bibitem [{\citenamefont {Carter-Fenk}\ and\ \citenamefont
  {Herbert}(2020)}]{carter2020state}%
  \BibitemOpen
  \bibfield  {author} {\bibinfo {author} {\bibfnamefont {K.}~\bibnamefont
  {Carter-Fenk}}\ and\ \bibinfo {author} {\bibfnamefont {J.~M.}\ \bibnamefont
  {Herbert}},\ }\bibfield  {title} {\enquote {\bibinfo {title} {State-targeted
  energy projection: A simple and robust approach to orbital relaxation of
  non-aufbau self-consistent field solutions},}\ }\href@noop {} {\bibfield
  {journal} {\bibinfo  {journal} {Journal of Chemical Theory and Computation}\
  }\textbf {\bibinfo {volume} {16}},\ \bibinfo {pages} {5067--5082} (\bibinfo
  {year} {2020})}\BibitemShut {NoStop}%
\bibitem [{\citenamefont {Helgaker}, \citenamefont {Jorgensen},\ and\
  \citenamefont {Olsen}(2013)}]{helgaker2013molecular}%
  \BibitemOpen
  \bibfield  {author} {\bibinfo {author} {\bibfnamefont {T.}~\bibnamefont
  {Helgaker}}, \bibinfo {author} {\bibfnamefont {P.}~\bibnamefont {Jorgensen}},
  \ and\ \bibinfo {author} {\bibfnamefont {J.}~\bibnamefont {Olsen}},\
  }\href@noop {} {\emph {\bibinfo {title} {Molecular electronic-structure
  theory}}}\ (\bibinfo  {publisher} {John Wiley \& Sons},\ \bibinfo {year}
  {2013})\BibitemShut {NoStop}%
\bibitem [{\citenamefont {L{\"o}wdin}(1950)}]{lowdin1950non}%
  \BibitemOpen
  \bibfield  {author} {\bibinfo {author} {\bibfnamefont {P.-O.}\ \bibnamefont
  {L{\"o}wdin}},\ }\bibfield  {title} {\enquote {\bibinfo {title} {On the
  non-orthogonality problem connected with the use of atomic wave functions in
  the theory of molecules and crystals},}\ }\href@noop {} {\bibfield  {journal}
  {\bibinfo  {journal} {The Journal of Chemical Physics}\ }\textbf {\bibinfo
  {volume} {18}},\ \bibinfo {pages} {365--375} (\bibinfo {year}
  {1950})}\BibitemShut {NoStop}%
\bibitem [{\citenamefont {Barrett}\ \emph {et~al.}(1994)\citenamefont
  {Barrett}, \citenamefont {Berry}, \citenamefont {Chan}, \citenamefont
  {Demmel}, \citenamefont {Donato}, \citenamefont {Dongarra}, \citenamefont
  {Eijkhout}, \citenamefont {Pozo}, \citenamefont {Romine},\ and\ \citenamefont
  {Van~der Vorst}}]{templates-linear-systems}%
  \BibitemOpen
  \bibfield  {author} {\bibinfo {author} {\bibfnamefont {R.}~\bibnamefont
  {Barrett}}, \bibinfo {author} {\bibfnamefont {M.}~\bibnamefont {Berry}},
  \bibinfo {author} {\bibfnamefont {T.~F.}\ \bibnamefont {Chan}}, \bibinfo
  {author} {\bibfnamefont {J.}~\bibnamefont {Demmel}}, \bibinfo {author}
  {\bibfnamefont {J.}~\bibnamefont {Donato}}, \bibinfo {author} {\bibfnamefont
  {J.}~\bibnamefont {Dongarra}}, \bibinfo {author} {\bibfnamefont
  {V.}~\bibnamefont {Eijkhout}}, \bibinfo {author} {\bibfnamefont
  {R.}~\bibnamefont {Pozo}}, \bibinfo {author} {\bibfnamefont {C.}~\bibnamefont
  {Romine}}, \ and\ \bibinfo {author} {\bibfnamefont {H.}~\bibnamefont {Van~der
  Vorst}},\ }\href@noop {} {\emph {\bibinfo {title} {Templates for the solution
  of linear systems: building blocks for iterative methods}}}\ (\bibinfo
  {publisher} {SIAM},\ \bibinfo {year} {1994})\BibitemShut {NoStop}%
\bibitem [{\citenamefont {Rubensson}\ and\ \citenamefont
  {Sa{\l}ek}(2005)}]{rubensson2005systematic}%
  \BibitemOpen
  \bibfield  {author} {\bibinfo {author} {\bibfnamefont {E.~H.}\ \bibnamefont
  {Rubensson}}\ and\ \bibinfo {author} {\bibfnamefont {P.}~\bibnamefont
  {Sa{\l}ek}},\ }\bibfield  {title} {\enquote {\bibinfo {title} {Systematic
  sparse matrix error control for linear scaling electronic structure
  calculations},}\ }\href@noop {} {\bibfield  {journal} {\bibinfo  {journal}
  {Journal of computational chemistry}\ }\textbf {\bibinfo {volume} {26}},\
  \bibinfo {pages} {1628--1637} (\bibinfo {year} {2005})}\BibitemShut {NoStop}%
\bibitem [{\citenamefont {Koch}, \citenamefont {S{\'a}nchez~de Mer{\'a}s},\
  and\ \citenamefont {Pedersen}(2003)}]{koch2003reduced}%
  \BibitemOpen
  \bibfield  {author} {\bibinfo {author} {\bibfnamefont {H.}~\bibnamefont
  {Koch}}, \bibinfo {author} {\bibfnamefont {A.}~\bibnamefont {S{\'a}nchez~de
  Mer{\'a}s}}, \ and\ \bibinfo {author} {\bibfnamefont {T.~B.}\ \bibnamefont
  {Pedersen}},\ }\bibfield  {title} {\enquote {\bibinfo {title} {Reduced
  scaling in electronic structure calculations using cholesky
  decompositions},}\ }\href@noop {} {\bibfield  {journal} {\bibinfo  {journal}
  {Journal of Chemical Physics}\ }\textbf {\bibinfo {volume} {118}},\ \bibinfo
  {pages} {9481--9484} (\bibinfo {year} {2003})}\BibitemShut {NoStop}%
\bibitem [{\citenamefont {Dunning~Jr}(1989)}]{dunning1989gaussian}%
  \BibitemOpen
  \bibfield  {author} {\bibinfo {author} {\bibfnamefont {T.~H.}\ \bibnamefont
  {Dunning~Jr}},\ }\bibfield  {title} {\enquote {\bibinfo {title} {Gaussian
  basis sets for use in correlated molecular calculations. i. the atoms boron
  through neon and hydrogen},}\ }\href@noop {} {\bibfield  {journal} {\bibinfo
  {journal} {The Journal of chemical physics}\ }\textbf {\bibinfo {volume}
  {90}},\ \bibinfo {pages} {1007--1023} (\bibinfo {year} {1989})}\BibitemShut
  {NoStop}%
\bibitem [{\citenamefont {Smith}\ \emph {et~al.}(2020)\citenamefont {Smith},
  \citenamefont {Burns}, \citenamefont {Simmonett}, \citenamefont {Parrish},
  \citenamefont {Schieber}, \citenamefont {Galvelis}, \citenamefont {Kraus},
  \citenamefont {Kruse}, \citenamefont {Remigio}, \citenamefont {Alenaizan},
  \citenamefont {James}, \citenamefont {Lehtola}, \citenamefont {Misiewicz},
  \citenamefont {Scheurer}, \citenamefont {Shaw}, \citenamefont {Schriber},
  \citenamefont {Xie}, \citenamefont {Glick}, \citenamefont {Sirianni},
  \citenamefont {O’Brien}, \citenamefont {Waldrop}, \citenamefont {Kumar},
  \citenamefont {Hohenstein}, \citenamefont {Pritchard}, \citenamefont
  {Brooks}, \citenamefont {III}, \citenamefont {Sokolov}, \citenamefont
  {Patkowski}, \citenamefont {III}, \citenamefont {Bozkaya}, \citenamefont
  {King}, \citenamefont {Evangelista}, \citenamefont {Turney}, \citenamefont
  {Crawford},\ and\ \citenamefont {Sherrill}}]{psi4}%
  \BibitemOpen
  \bibfield  {author} {\bibinfo {author} {\bibfnamefont {D.~G.~A.}\
  \bibnamefont {Smith}}, \bibinfo {author} {\bibfnamefont {L.~A.}\ \bibnamefont
  {Burns}}, \bibinfo {author} {\bibfnamefont {A.~C.}\ \bibnamefont
  {Simmonett}}, \bibinfo {author} {\bibfnamefont {R.~M.}\ \bibnamefont
  {Parrish}}, \bibinfo {author} {\bibfnamefont {M.~C.}\ \bibnamefont
  {Schieber}}, \bibinfo {author} {\bibfnamefont {R.}~\bibnamefont {Galvelis}},
  \bibinfo {author} {\bibfnamefont {P.}~\bibnamefont {Kraus}}, \bibinfo
  {author} {\bibfnamefont {H.}~\bibnamefont {Kruse}}, \bibinfo {author}
  {\bibfnamefont {R.~D.}\ \bibnamefont {Remigio}}, \bibinfo {author}
  {\bibfnamefont {A.}~\bibnamefont {Alenaizan}}, \bibinfo {author}
  {\bibfnamefont {A.~M.}\ \bibnamefont {James}}, \bibinfo {author}
  {\bibfnamefont {S.}~\bibnamefont {Lehtola}}, \bibinfo {author} {\bibfnamefont
  {J.~P.}\ \bibnamefont {Misiewicz}}, \bibinfo {author} {\bibfnamefont
  {M.}~\bibnamefont {Scheurer}}, \bibinfo {author} {\bibfnamefont {R.~A.}\
  \bibnamefont {Shaw}}, \bibinfo {author} {\bibfnamefont {J.~B.}\ \bibnamefont
  {Schriber}}, \bibinfo {author} {\bibfnamefont {Y.}~\bibnamefont {Xie}},
  \bibinfo {author} {\bibfnamefont {Z.~L.}\ \bibnamefont {Glick}}, \bibinfo
  {author} {\bibfnamefont {D.~A.}\ \bibnamefont {Sirianni}}, \bibinfo {author}
  {\bibfnamefont {J.~S.}\ \bibnamefont {O’Brien}}, \bibinfo {author}
  {\bibfnamefont {J.~M.}\ \bibnamefont {Waldrop}}, \bibinfo {author}
  {\bibfnamefont {A.}~\bibnamefont {Kumar}}, \bibinfo {author} {\bibfnamefont
  {E.~G.}\ \bibnamefont {Hohenstein}}, \bibinfo {author} {\bibfnamefont
  {B.~P.}\ \bibnamefont {Pritchard}}, \bibinfo {author} {\bibfnamefont {B.~R.}\
  \bibnamefont {Brooks}}, \bibinfo {author} {\bibfnamefont {H.~F.~S.}\
  \bibnamefont {III}}, \bibinfo {author} {\bibfnamefont {A.~Y.}\ \bibnamefont
  {Sokolov}}, \bibinfo {author} {\bibfnamefont {K.}~\bibnamefont {Patkowski}},
  \bibinfo {author} {\bibfnamefont {A.~E.~D.}\ \bibnamefont {III}}, \bibinfo
  {author} {\bibfnamefont {U.}~\bibnamefont {Bozkaya}}, \bibinfo {author}
  {\bibfnamefont {R.~A.}\ \bibnamefont {King}}, \bibinfo {author}
  {\bibfnamefont {F.~A.}\ \bibnamefont {Evangelista}}, \bibinfo {author}
  {\bibfnamefont {J.~M.}\ \bibnamefont {Turney}}, \bibinfo {author}
  {\bibfnamefont {T.~D.}\ \bibnamefont {Crawford}}, \ and\ \bibinfo {author}
  {\bibfnamefont {C.~D.}\ \bibnamefont {Sherrill}},\ }\bibfield  {title}
  {\enquote {\bibinfo {title} {Psi4 1.4: Open-source software for
  high-throughput quantum chemistry},}\ }\href@noop {} {\bibfield  {journal}
  {\bibinfo  {journal} {J. Chem. Phys.}\ }\textbf {\bibinfo {volume} {152}},\
  \bibinfo {pages} {184108} (\bibinfo {year} {2020})}\BibitemShut {NoStop}%
\bibitem [{\citenamefont {Valeev}(2025)}]{Libint2}%
  \BibitemOpen
  \bibfield  {author} {\bibinfo {author} {\bibfnamefont {E.~F.}\ \bibnamefont
  {Valeev}},\ }\bibfield  {title} {\enquote {\bibinfo {title} {Libint: A
  library for the evaluation of molecular integrals of many-body operators over
  gaussian functions, version 2.8.2},}\ }\href@noop {} {\bibfield  {journal}
  {\bibinfo  {journal} {http://libint.valeyev.net}\ } (\bibinfo {year}
  {2025})}\BibitemShut {NoStop}%
\bibitem [{\citenamefont {Pipek}\ and\ \citenamefont
  {Mezey}(1989)}]{pipek1989fast}%
  \BibitemOpen
  \bibfield  {author} {\bibinfo {author} {\bibfnamefont {J.}~\bibnamefont
  {Pipek}}\ and\ \bibinfo {author} {\bibfnamefont {P.~G.}\ \bibnamefont
  {Mezey}},\ }\bibfield  {title} {\enquote {\bibinfo {title} {A fast intrinsic
  localization procedure applicable for ab initio and semiempirical linear
  combination of atomic orbital wave functions},}\ }\href@noop {} {\bibfield
  {journal} {\bibinfo  {journal} {The Journal of Chemical Physics}\ }\textbf
  {\bibinfo {volume} {90}},\ \bibinfo {pages} {4916--4926} (\bibinfo {year}
  {1989})}\BibitemShut {NoStop}%
\bibitem [{\citenamefont {Clune}\ \emph {et~al.}(2023)\citenamefont {Clune},
  \citenamefont {Shea}, \citenamefont {Hardikar}, \citenamefont {Tuckman},\
  and\ \citenamefont {Neuscamman}}]{clune2023thiel}%
  \BibitemOpen
  \bibfield  {author} {\bibinfo {author} {\bibfnamefont {R.}~\bibnamefont
  {Clune}}, \bibinfo {author} {\bibfnamefont {J.~A.}\ \bibnamefont {Shea}},
  \bibinfo {author} {\bibfnamefont {T.~S.}\ \bibnamefont {Hardikar}}, \bibinfo
  {author} {\bibfnamefont {H.}~\bibnamefont {Tuckman}}, \ and\ \bibinfo
  {author} {\bibfnamefont {E.}~\bibnamefont {Neuscamman}},\ }\bibfield  {title}
  {\enquote {\bibinfo {title} {Studying excited-state-specific perturbation
  theory on the thiel set},}\ }\href@noop {} {\bibfield  {journal} {\bibinfo
  {journal} {The Journal of Chemical Physics}\ }\textbf {\bibinfo {volume}
  {158}},\ \bibinfo {pages} {224113} (\bibinfo {year} {2023})}\BibitemShut
  {NoStop}%
\bibitem [{\citenamefont {Schreiber}\ \emph {et~al.}(2008)\citenamefont
  {Schreiber}, \citenamefont {Silva-Junior}, \citenamefont {Sauer},\ and\
  \citenamefont {Thiel}}]{thiel2008benchmark}%
  \BibitemOpen
  \bibfield  {author} {\bibinfo {author} {\bibfnamefont {M.}~\bibnamefont
  {Schreiber}}, \bibinfo {author} {\bibfnamefont {M.~R.}\ \bibnamefont
  {Silva-Junior}}, \bibinfo {author} {\bibfnamefont {S.}~\bibnamefont {Sauer}},
  \ and\ \bibinfo {author} {\bibfnamefont {W.}~\bibnamefont {Thiel}},\
  }\bibfield  {title} {\enquote {\bibinfo {title} {Benchmarks for
  electronically excited states: Caspt2, cc2, ccsd, and cc3},}\ }\href@noop {}
  {\bibfield  {journal} {\bibinfo  {journal} {The Journal of chemical physics}\
  }\textbf {\bibinfo {volume} {128}},\ \bibinfo {pages} {134110} (\bibinfo
  {year} {2008})}\BibitemShut {NoStop}%
\bibitem [{\citenamefont {Mester}\ and\ \citenamefont
  {K{\'a}llay}(2022)}]{mester2022charge}%
  \BibitemOpen
  \bibfield  {author} {\bibinfo {author} {\bibfnamefont {D.}~\bibnamefont
  {Mester}}\ and\ \bibinfo {author} {\bibfnamefont {M.}~\bibnamefont
  {K{\'a}llay}},\ }\bibfield  {title} {\enquote {\bibinfo {title}
  {Charge-transfer excitations within density functional theory: How accurate
  are the most recommended approaches?}}\ }\href@noop {} {\bibfield  {journal}
  {\bibinfo  {journal} {Journal of Chemical Theory and Computation}\ }\textbf
  {\bibinfo {volume} {18}},\ \bibinfo {pages} {1646--1662} (\bibinfo {year}
  {2022})}\BibitemShut {NoStop}%
\end{thebibliography}
%\printbibliography

%Note: revtex disables \onecolumn, but can use this instead:
\onecolumngrid

\begin{center}

$\vspace{10mm}$\\

{\huge Supplemental Information}

\end{center}

\noindent
\rule[8mm]{0mm}{0mm}{\Large Geometries in Angstroms}

\noindent
\rule[8mm]{0mm}{0mm}Formaldehyde

H       0.000000        0.934473       -0.588078

H       0.000000       -0.934473       -0.588078

C       0.000000        0.000000        0.000000

O       0.000000        0.000000        1.221104

\vspace{1mm}

\noindent
\rule[8mm]{0mm}{0mm}Acetone

H  0.000000  2.136732 -0.112445

H  0.000000 -2.136732 -0.112445

H -0.881334  1.333733 -1.443842

H  0.881334 -1.333733 -1.443842

H -0.881334 -1.333733 -1.443842

H  0.881334  1.333733 -1.443842

C  0.000000  0.000000  0.000000

C  0.000000  1.287253 -0.795902

C  0.000000 -1.287253 -0.795902

O  0.000000  0.000000  1.227600

\noindent
\rule[8mm]{0mm}{0mm}Formamide

H   -0.927427    -0.600301     0.000000

H    1.070498    -1.782390     0.000000

H    2.024514    -0.325050     0.000000

C    0.000000     0.000000     0.000000

O    0.000000     1.225060     0.000000

N    1.119392    -0.775069     0.000000

\noindent
\rule[8mm]{0mm}{0mm}Acetamide

H  1.173209 -1.735763  0.000000

H  2.035841 -0.226201  0.000000

H -2.121189 -0.156089  0.000000

H -1.310647 -1.472742  0.885504

H -1.310647 -1.472742 -0.885504

C  0.000000  0.000000  0.000000

C -1.267042 -0.831610  0.000000

O  0.000000  1.229439  0.000000

N  1.158967 -0.727718  0.000000

\clearpage

\noindent
\rule[8mm]{0mm}{0mm}Propanamide

H 1.171887 -1.734653 0.000000

H 2.036508 -0.225526 0.000000

H -1.256737 -1.492368 0.877197

H -1.256737 -1.492368 -0.877197

H -3.420939 -0.590421 0.000000

H -2.544313 0.678541 -0.880209

H -2.544313 0.678541 0.880209

C 0.000000 0.000000 0.000000

C -1.272727 -0.833216 0.000000

C -2.523376 0.033790 0.000000

O 0.000000 1.230373 0.000000

N 1.159100 -0.726409 0.000000

\noindent
\rule[8mm]{0mm}{0mm}Butyramide

 O          2.7421955470       -1.1684209706        0.0000000000
 
 N          1.7333629654        0.8737403490        0.0000000000
 
 H          1.8135394062        1.8800953350        0.0000000000
 
 H          0.8147524941        0.4485619258        0.0000000000
 
 C          2.8317665061        0.0538644993        0.0000000000
 
 C          4.1657929824        0.7937027253        0.0000000000
 
 H          4.1999078211        1.4575686986        0.8837151630
 
 H          4.1999078211        1.4575686986       -0.8837151630
 
 C          5.3624745950       -0.1544386256        0.0000000000
 
 H          5.2777086892       -0.8174003572        0.8750630033
 
 H          5.2777086892       -0.8174003572       -0.8750630033
 
 C          6.7126179310        0.5926734420        0.0000000000
 
 H          7.3173349939        0.3452264815        0.8860809571
 
 H          7.3173349939        0.3452264815       -0.8860809571
 
 H          6.5724945642        1.6863316742        0.0000000000
 
\noindent
\rule[8mm]{0mm}{0mm}Imidazole

H   0.000000  2.119822  0.714354

H   0.000000  1.202262 -1.904898

H   0.000000 -2.104815  0.663782

H   0.000000 -0.010302  2.116597

C   0.000000  1.120107  0.305897

C   0.000000  0.635508 -0.983749

C   0.000000 -1.091835  0.283881

N   0.000000 -0.741378 -0.994001

N   0.000000  0.000000  1.104571

\noindent
\rule[8mm]{0mm}{0mm}Uracil

H  -2.025413 -1.517742 0.000000

H  -0.021861  1.995767 0.000000

H   2.182391 -1.602586 0.000000

H  -0.026659 -2.791719 0.000000

C  -1.239290  0.359825 0.000000

C   1.279718  0.392094 0.000000

C   1.243729 -1.064577 0.000000

C   0.055755 -1.709579 0.000000

O  -2.308803  0.954763 0.000000

O   2.287387  1.092936 0.000000

N  -1.139515 -1.026364 0.000000

N   0.000000  0.978951 0.000000

\clearpage

\noindent
\rule[8mm]{0mm}{0mm}Butadiene

H 1.080977 -2.558832 0.000000

H -1.080977 2.558832 0.000000

H 2.103773 -1.017723 0.000000

H -2.103773 1.017723 0.000000

H -0.973565 -1.219040 0.000000

H 0.973565 1.219040 0.000000

C 0.000000 0.728881 0.000000

C 0.000000 -0.728881 0.000000

C 1.117962 -1.474815 0.000000

C -1.117962 1.474815 0.000000

\noindent
\rule[8mm]{0mm}{0mm}Hexatriene

H -0.953777 1.207691 0.000000

H 0.953777 -1.207691 0.000000

H 2.155816 0.952317 0.000000

H -2.155816 -0.952317 0.000000

H 2.125769 3.402692 0.000000

H -2.125769 -3.402692 0.000000

H 0.275642 3.397162 0.000000

H -0.275642 -3.397162 0.000000

C 0.000000 0.676808 0.000000

C 0.000000 -0.676808 0.000000

C 1.204938 1.485654 0.000000

C -1.204938 -1.485654 0.000000

C 1.203567 2.831663 0.000000

C -1.203567 -2.831663 0.000000

\noindent
\rule[8mm]{0mm}{0mm}Cyclopentadiene

H -0.879859 0.000000 1.874608

H 0.879859 0.000000 1.874608

H 0.000000 2.211693 0.612518

H 0.000000 -2.211693 0.612518

H 0.000000 1.349811 -1.886050

H 0.000000 -1.349811 -1.886050

C 0.000000 0.000000 1.215652

C 0.000000 -1.177731 0.285415

C 0.000000 1.177731 0.285415

C 0.000000 -0.732372 -0.993420

C 0.000000 0.732372 -0.993420

\noindent
\rule[8mm]{0mm}{0mm}Furan

H   0.000000  2.051058  0.851533

H   0.000000 -2.051058  0.851533

H   0.000000  1.371979 -1.821224

H   0.000000 -1.371979 -1.821224

C   0.000000  1.095840  0.348301

C   0.000000 -1.095840  0.348301

C   0.000000  0.714027 -0.963274

C   0.000000 -0.714027 -0.963274

O   0.000000  0.000000  1.164881

\clearpage

\noindent
\rule[8mm]{0mm}{0mm}Pyridine 

H   0.000000  2.061947  1.308539

H   0.000000 -2.061947  1.308539

H   0.000000  2.156804 -1.184054

H   0.000000 -2.156804 -1.184054

H   0.000000  0.000000 -2.475074

C   0.000000  1.145417  0.721005

C   0.000000 -1.145417  0.721005

C   0.000000  1.197637 -0.673735

C   0.000000 -1.197637 -0.673735

C   0.000000  0.000000 -1.387901

N   0.000000  0.000000  1.426610

\noindent
\rule[8mm]{0mm}{0mm}NH$_2$(CH$_2$)$_3$CHCFCHCHF

N      -4.9688269403     0.1239667099     0.0000000000
      
H      -5.0613859244    -0.4991562014    -0.8079465758
      
H      -5.0613859244    -0.4991562013     0.8079465758
      
C      -3.6057745448     0.6645961101     0.0000000000
      
H      -3.4993620373     1.3199109377    -0.8841426204
      
H      -3.4993620372     1.3199109378     0.8841426204
      
C      -2.4766960517    -0.3715501691     0.0000000000
      
H      -2.5822355862    -1.0251932334    -0.8875678434
      
H      -2.5822355862    -1.0251932334     0.8875678434
      
C      -1.0797781746     0.2582150590     0.0000000000
      
H      -0.9843970276     0.9167940360    -0.8853740804
      
H      -0.9843970275     0.9167940360     0.8853740804
      
C       0.0122205125    -0.7871949315     0.0000000000
      
H      -0.2695867430    -1.8468047378     0.0000000000
      
C       1.3373243636    -0.5161927473     0.0000000000
      
F       2.2212995508    -1.5419346784     0.0000000000
      
C       1.9249025849     0.8171883074     0.0000000000
      
H       1.2327018752     1.6625928602     0.0000000000
      
C       3.2355612968     1.1388835824     0.0000000000
      
F       4.2336619211     0.2467147306     0.0000000000
      
H       3.5911368699     2.1736107961     0.0000000000

\clearpage

\noindent
\rule[8mm]{0mm}{0mm}NH$_2$(CH$_2$)$_4$CHCFCHCHF

N      -5.5865626788     0.5504335500     0.0000000000
      
H      -5.5141561052     1.1763740727    -0.8078614771
      
H      -5.5141561052     1.1763740727     0.8078614771
      
C      -4.4102455166    -0.3255427299     0.0000000000
      
H      -4.4775791210    -0.9859657440    -0.8839531546
      
H      -4.4775791210    -0.9859657440     0.8839531546
      
C      -3.0486584189     0.3805167861     0.0000000000
      
H      -2.9853748374     1.0411284301    -0.8872983801
      
H      -2.9853748374     1.0411284301     0.8872983801
      
C      -1.8647362775    -0.5879130344     0.0000000000
      
H      -1.9265714930    -1.2478775858    -0.8862841399
      
H      -1.9265714930    -1.2478775858     0.8862841399
      
C      -0.5061089222     0.1218329181     0.0000000000
      
H      -0.4493346165     0.7843740780    -0.8855455072
      
H      -0.4493346165     0.7843740780     0.8855455072
      
C       0.6428788694    -0.8605903979     0.0000000000
      
H       0.4196019212    -1.9341854976     0.0000000000
      
C       1.9511789852    -0.5178947057     0.0000000000
      
F       2.8901148457    -1.4936367305     0.0000000000
      
C       2.4651835000     0.8455736220     0.0000000000
      
H       1.7275366013     1.6516771607     0.0000000000
      
C       3.7560850727     1.2389530542     0.0000000000
      
F       4.8020396122     0.4033612066     0.0000000000
      
H       4.0540326815     2.2917293747     0.0000000000

\noindent
\rule[8mm]{0mm}{0mm}NH$_2$(CH$_2$)$_5$CHCFCHCHF

N      -6.3198309586     0.0626680830     0.0000000000
      
H      -6.3777970675    -0.5652269150    -0.8075624426
      
H      -6.3777970675    -0.5652269150     0.8075624426
      
C      -4.9865709269     0.6746559978     0.0000000000
      
H      -4.9155888811     1.3348681709    -0.8839765926
      
H      -4.9155888810     1.3348681709     0.8839765926
      
C      -3.8019925138    -0.2990868853     0.0000000000
      
H      -3.8775381182    -0.9585112270    -0.8870689014
      
H      -3.8775381182    -0.9585112270     0.8870689014
      
C      -2.4401836629     0.4001341203     0.0000000000
      
H      -2.3673684256     1.0606025299    -0.8859642071
      
H      -2.3673684256     1.0606025299     0.8859642071
      
C      -1.2611372025    -0.5749520761     0.0000000000
      
H      -1.3269179515    -1.2341116360    -0.8865024692
      
H      -1.3269179515    -1.2341116360     0.8865024692
      
C       0.1000968718     0.1296121986     0.0000000000
      
H       0.1591972519     0.7920907674    -0.8854962960
      
H       0.1591972519     0.7920907674     0.8854962960
      
C       1.2458721593    -0.8565972580     0.0000000000
      
H       1.0192295686    -1.9295034734     0.0000000000
      
C       2.5552722394    -0.5180426117     0.0000000000
      
F       3.4911522608    -1.4968794791     0.0000000000
      
C       3.0735036580     0.8438002452     0.0000000000
      
H       2.3383068409     1.6521349390     0.0000000000
      
C       4.3656142968     1.2331260639     0.0000000000
      
F       5.4089670151     0.3941601223     0.0000000000
      
H       4.6668906497     2.2849486301     0.0000000000

\clearpage

\noindent
\rule[8mm]{0mm}{0mm}NH$_2$(CH$_2$)$_6$CHCFCHCHF

N      -6.9428861880     0.5386334927     0.0000000000
      
H      -6.8873737135     1.1668284925    -0.8075421244
      
H      -6.8873737135     1.1668284924     0.8075421244
      
C      -5.7408600520    -0.3025153712     0.0000000000
      
H      -5.7891528254    -0.9648085634    -0.8839073487
      
H      -5.7891528254    -0.9648085635     0.8839073487
      
C      -4.4007000146     0.4429732380     0.0000000000
      
H      -4.3572796523     1.1054791339    -0.8870902452
      
H      -4.3572796523     1.1054791339     0.8870902452
      
C      -3.1863105997    -0.4889349077     0.0000000000
      
H      -3.2332392399    -1.1517324907    -0.8857660048
      
H      -3.2332392399    -1.1517324907     0.8857660048
      
C      -1.8491203665     0.2572845264     0.0000000000
      
H      -1.7997662655     0.9193704803    -0.8862395602
      
H      -1.7997662656     0.9193704803     0.8862395602
      
C      -0.6384973661    -0.6782693954     0.0000000000
      
H      -0.6826350341    -1.3392425991    -0.8865006344
      
H      -0.6826350341    -1.3392425991     0.8865006344
      
C       0.6991621829     0.0703013763     0.0000000000
      
H       0.7368306453     0.7342997869    -0.8855308084
      
H       0.7368306453     0.7342997869     0.8855308084
      
C       1.8763741202    -0.8781888687     0.0000000000
      
H       1.6844971099    -1.9578522280     0.0000000000
      
C       3.1741266654    -0.4974602421     0.0000000000
      
F       4.1412737308    -1.4454426194     0.0000000000
      
C       3.6480922589     0.8804359795     0.0000000000
      
H       2.8870516752     1.6644967797     0.0000000000
      
C       4.9268872637     1.3114926815     0.0000000000
      
F       5.9970028526     0.5069014749     0.0000000000
      
H       5.1938706443     2.3725421329     0.0000000000

\noindent
\rule[8mm]{0mm}{0mm}Water flyby geometry 0
 
C          0.0000000000        0.5322959226       -0.5296478177
 
C          0.0000000000        1.3276317456        0.5028067487
 
H          0.0000000000        2.3780628748        0.2805126514
 
C         -0.0000000000        0.8819629219        1.9505408478
 
O          0.0000000000       -0.1717315537       -1.4656264881
 
H          0.8859325794        1.2652724975        2.4438870471
 
H         -0.8859325794        1.2652724975        2.4438870471
 
C         -0.0000000000       -0.6486155553        2.1238248145
 
H          0.8809524780       -1.0781856451        1.6605910638
 
H         -0.8809524780       -1.0781856451        1.6605910638
 
N         -0.0000000000       -1.0815456784        3.5230902183
 
H         -0.8024966813       -0.7519465777        4.0253300732
 
H          0.8024966813       -0.7519465777        4.0253300732
 
O          0.0000000000       -0.4269441581       -5.1013266519
 
H          0.0000000000       -0.3126962645       -4.1562083678
 
H          0.0000000000       -1.3473665138       -5.3342704574

\clearpage

\noindent
\rule[8mm]{0mm}{0mm}Water flyby geometry 1
 
C          0.0000000000        0.7250983386       -0.5727012201
 
C          0.0000000000        1.4210962889        0.5291646321
 
H          0.0000000000        2.4876340238        0.4054008126
 
C         -0.0000000000        0.8428814187        1.9292440176
 
O          0.0000000000        0.1110528342       -1.5700270832
 
H          0.8859325794        1.1787094400        2.4560611047
 
H         -0.8859325794        1.1787094400        2.4560611047
 
C         -0.0000000000       -0.6971756088        1.9596111119
 
H          0.8809524780       -1.0818611470        1.4584793862
 
H         -0.8809524780       -1.0818611470        1.4584793862
 
N         -0.0000000000       -1.2582048364        3.3126145737
 
H         -0.8024966813       -0.9766811783        3.8432979717
 
H          0.8024966813       -0.9766811783        3.8432979717
 
O          0.0000000000       -0.6537070850       -4.6842832417
 
H          0.0000000000       -0.3385147139       -3.7859444727
 
H          0.0000000000       -1.6023988062       -4.7137658599

\noindent
\rule[8mm]{0mm}{0mm}Water flyby geometry 2
 
C          0.0000000000        0.9348485119       -0.5717090481
 
C          0.0000000000        1.5109936775        0.5972983305
 
H          0.0000000000        2.5846339294        0.5864894254
 
C         -0.0000000000        0.7886141086        1.9287345778
 
O          0.0000000000        0.4291962667       -1.6281306332
 
H          0.8859325794        1.0671218686        2.4879752511
 
H         -0.8859325794        1.0671218686        2.4879752511
 
C         -0.0000000000       -0.7460834917        1.7968214094
 
H          0.8809524780       -1.0758810619        1.2579804936
 
H         -0.8809524780       -1.0758810619        1.2579804936
 
N         -0.0000000000       -1.4464175256        3.0832522870
 
H         -0.8024966813       -1.2223194152        3.6406215326
 
H          0.8024966813       -1.2223194152        3.6406215326
 
O          0.0000000000       -0.9028670956       -4.3683963783
 
H          0.0000000000       -0.3380839377       -3.6030477918
 
H          0.0000000000       -1.8181054560       -4.1176111018

\noindent
\rule[8mm]{0mm}{0mm}Water flyby geometry 3
 
C          0.0000000000        1.1611244863       -0.5062819371
 
C          0.0000000000        1.5852617092        0.7260447725
 
H          0.0000000000        2.6516949665        0.8507056352
 
C         -0.0000000000        0.7007573392        1.9557629278
 
O          0.0000000000        0.7927206961       -1.6180325413
 
H          0.8859325794        0.9065230823        2.5456588187
 
H         -0.8859325794        0.9065230823        2.5456588187
 
C         -0.0000000000       -0.8050561164        1.6313809845
 
H          0.8809524780       -1.0642746271        1.0552544936
 
H         -0.8809524780       -1.0642746271        1.0552544936
 
N         -0.0000000000       -1.6620158521        2.8192329000
 
H         -0.8024966813       -1.5099894694        3.4004113735
 
H          0.8024966813       -1.5099894694        3.4004113735
 
O          0.0000000000       -1.1304877945       -4.1841031281
 
H          0.0000000000       -0.3902021755       -3.5857012208
 
H          0.0000000000       -1.9526627788       -3.7098794917

\clearpage

\noindent
\rule[8mm]{0mm}{0mm}Water flyby geometry 4
 
C          0.0000000000        1.3971259839       -0.3392139539
 
C          0.0000000000        1.6151585490        0.9456919974
 
H          0.0000000000        2.6471105632        1.2421633506
 
C         -0.0000000000        0.5424030563        2.0151546544
 
O          0.0000000000        1.2144711153       -1.4960838505
 
H          0.8859325794        0.6494731985        2.6306647581
 
H         -0.8859325794        0.6494731985        2.6306647581
 
C         -0.0000000000       -0.8905897243        1.4501506729
 
H          0.8809524780       -1.0526405125        0.8395317149
 
H         -0.8809524780       -1.0526405125        0.8395317149
 
N         -0.0000000000       -1.9293572880        2.4827852211
 
H         -0.8024966813       -1.8738927615        3.0809525266
 
H          0.8024966813       -1.8738927615        3.0809525266
 
O          0.0000000000       -1.2525166054       -4.1788847639
 
H          0.0000000000       -0.4300344346       -3.6997932456
 
H          0.0000000000       -1.9944099482       -3.5864354471

\noindent
\rule[8mm]{0mm}{0mm}Water flyby geometry 5
 
C          0.0000000000        1.6088977538       -0.0157637105
 
C          0.0000000000        1.5409866247        1.2857391061
 
H          0.0000000000        2.4832548302        1.8004728740
 
C         -0.0000000000        0.2605222095        2.0950385583
 
O          0.0000000000        1.6833636757       -1.1845945730
 
H          0.8859325794        0.2305529745        2.7190726504
 
H         -0.8859325794        0.2305529745        2.7190726504
 
C         -0.0000000000       -1.0144425906        1.2306528997
 
H          0.8809524780       -1.0391946427        0.5993817172
 
H         -0.8809524780       -1.0391946427        0.5993817172
 
N         -0.0000000000       -2.2536950650        2.0114380598
 
H         -0.8024966813       -2.3302352244        2.6072753220
 
H          0.8024966813       -2.3302352244        2.6072753220
 
O          0.0000000000       -1.1363511129       -4.3990397297
 
H          0.0000000000       -0.2839610908       -3.9776708209
 
H          0.0000000000       -1.8375326766       -3.7585971946

\noindent
\rule[8mm]{0mm}{0mm}Water flyby geometry 6
 
C          0.0000000000        1.7181445710        0.4663321449
 
C          0.0000000000        1.3083588788        1.7035052085
 
H          0.0000000000        2.0808942372        2.4491655487
 
C         -0.0000000000       -0.1405786888        2.1452456834
 
O          0.0000000000        2.0991550920       -0.6411613051
 
H          0.8859325794       -0.3345595332        2.7391212050
 
H         -0.8859325794       -0.3345595332        2.7391212050
 
C         -0.0000000000       -1.1414632571        0.9743793848
 
H          0.8809524780       -0.9983398362        0.3590488588
 
H         -0.8809524780       -0.9983398362        0.3590488588
 
N         -0.0000000000       -2.5431139759        1.3995233887
 
H         -0.8024966813       -2.7745476330        1.9538868989
 
H          0.8024966813       -2.7745476330        1.9538868989
 
O          0.0000000000       -0.7173317581       -4.8065965738
 
H          0.0000000000        0.1487317083       -4.4157611156
 
H          0.0000000000       -1.3965570940       -4.1428162802

\clearpage

\noindent
\rule[8mm]{0mm}{0mm}Water flyby geometry 7
 
C          0.0000000000        1.6965086842        0.9677171042
 
C          0.0000000000        0.9961527052        2.0668181031
 
H          0.0000000000        1.5624967081        2.9789993862
 
C         -0.0000000000       -0.5168489131        2.1401722091
 
O          0.0000000000        2.3372001598       -0.0127038217
 
H          0.8859325794       -0.8503941993        2.6684375261
 
H         -0.8859325794       -0.8503941993        2.6684375261
 
C         -0.0000000000       -1.2004277780        0.7598035106
 
H          0.8809524780       -0.9109349174        0.1982789310
 
H         -0.8809524780       -0.9109349174        0.1982789310
 
N         -0.0000000000       -2.6635177670        0.8286500484
 
H         -0.8024966813       -3.0236960550        1.3094327850
 
H          0.8024966813       -3.0236960550        1.3094327850
 
O          0.0000000000       -0.2135275056       -5.2831153361
 
H          0.0000000000        0.6241501393       -4.8351424906
 
H          0.0000000000       -0.9362778022       -4.6668146883

\noindent
\rule[8mm]{0mm}{0mm}Water flyby geometry 8
 
C          0.0000000000        1.6224167052        1.3748078432
 
C          0.0000000000        0.7368542554        2.3309985410
 
H          0.0000000000        1.1310019159        3.3297316608
 
C         -0.0000000000       -0.7648872938        2.1326880734
 
O          0.0000000000        2.4280591022        0.5247189732
 
H          0.8859325794       -1.1874985582        2.5928146020
 
H         -0.8859325794       -1.1874985582        2.5928146020
 
C         -0.0000000000       -1.1906822469        0.6523514677
 
H          0.8809524780       -0.8054679069        0.1516261117
 
H         -0.8809524780       -0.8054679069        0.1516261117
 
N         -0.0000000000       -2.6425103913        0.4585288664
 
H         -0.8024966813       -3.0828370391        0.8671764924
 
H          0.8024966813       -3.0828370391        0.8671764924
 
O          0.0000000000        0.1306492986       -5.8002598686
 
H          0.0000000000        0.9417008845       -5.3059163876
 
H          0.0000000000       -0.6257871868       -5.2257564341

\end{document}